\documentclass[twocolumn,twocolappendix]{aastex631}

\providecommand{\sorthelp}[1]{}
 
\usepackage{graphicx}
\usepackage{amssymb}
\usepackage{epstopdf}
\usepackage{graphicx}
\usepackage{bm}
\usepackage{xcolor} 
\usepackage{soul}
\usepackage[caption=false]{subfig}
\usepackage[version=4]{mhchem}

\newcommand{\be}{\begin{eqnarray}}

\newcommand{\ee}{\end{eqnarray}}

\usepackage[normalem]{ulem}

\providecommand{\sorthelp}[1]{}
\begin{document}

\preprint{APS/123-Q ED}

\title{Quijote PNG: The information content of the halo power spectrum and bispectrum}

\newcommand{\ens}{Laboratoire de Physique de l'\'{E}cole Normale Sup\'{e}rieure, ENS, Universite PSL, CNRS, Sorbonne Universit\'{e}, Universit\'{e} de Paris, F-75005 Paris, France}
\newcommand{\cnrs}{Sorbonne Universit\'{e}, CNRS, UMR 7095, Institut d'Astrophysique de Paris, 98 bis bd Arago, 75014 Paris, France}
\newcommand{\cca}{Center for Computational Astrophysics, Flatiron Institute, 162 5th Avenue, New York, NY 10010, USA}
\newcommand{\bologna}{Dipartimento di Fisica e Astronomia, Alma Mater Studiorum - University of Bologna, Via Piero Gobetti 93/2, 40129 Bologna BO, Italy}
\newcommand{\inaf}{INAF - Osservatorio Astronomico di Bologna, Via Piero Gobetti 93/3, 40129 Bologna BO, Italy}
\newcommand{\infn}{INFN - Istituto Nazionale di Fisica Nucleare, Sezione di Bologna, Viale Berti Pichat 6/2, 40127 Bologna BO, Italy
}
\newcommand{\infnPad}{INFN, Sezione di Padova, via Marzolo 8, I-35131, Padova, Italy}
\newcommand{\mpa}{Max-Planck-Institut f\"ur Astrophysik, Karl-Schwarzschild-Straße 1, 85748 Garching, Germany}
\newcommand{\ICC}{ICC, University of Barcelona, IEEC-UB, Martí i Franquès, 1, E-08028 Barcelona, Spain}
\newcommand{\ICREA}{ICREA, Pg. Lluís Companys 23, Barcelona, E-08010, Spain}
\newcommand{\Galilei}{Dipartimento di Fisica e Astronomia “G. Galilei”,Università degli Studi di Padova, via Marzolo 8, I-35131, Padova, Italy}
\newcommand{\uwc}{Department of Physics and Astronomy, University of the Western Cape, Cape Town 7535, South Africa}
\newcommand{\princeton}{Department of Astrophysical Sciences, Princeton University, 4 Ivy Lane, Princeton, NJ 08544 USA}
\author{William R Coulton}
\affiliation{\cca}
\author{ Francisco Villaescusa-Navarro}
\affiliation{\cca}
\affiliation{\princeton}
\author{Drew Jamieson}
\affiliation{\mpa}
\author{Marco Baldi}
\affiliation{\bologna}
\affiliation{\inaf}
\affiliation{\infn}
\author{Gabriel Jung}
\affiliation{\Galilei}
\affiliation{\infnPad}
\author{Dionysios Karagiannis}
\affiliation{\uwc}
\author{Michele Liguori}
\affiliation{\Galilei}
\affiliation{\infnPad}
\author{Licia Verde}
\affiliation{\ICREA} 
\affiliation{\ICC}
\author{Benjamin D. Wandelt}
\affiliation{\cnrs}
\affiliation{\cca}

\date{\today}% It is always \today, today,
    % but any date may be explicitly specified

\begin{abstract}
We investigate how much can be learnt about four types of primordial non-Gaussianity (PNG) from small-scale measurements of the halo field. Using the \textsc{quijote-png} simulations, we quantify the information content accessible with measurements of the halo power spectrum monopole and quadrupole, the matter power spectrum, the halo-matter cross spectrum and the halo bispectrum monopole. This analysis is the first to include small, non-linear scales, up to $k_\mathrm{max}=0.5 \mathrm{h/Mpc}$, and to explore whether these scales can break degeneracies with cosmological and nuisance parameters making use of thousands of N-body simulations. We perform all the halo measurements in redshift space with a single sample comprised of all halos with mass $>3.2 \times 10^{13}~h^{-1}M_\odot$.
For \emph{local} PNG, measurements of the scale dependent bias effect from the power spectrum using sample variance cancellation provide significantly tighter constraints than measurements of the halo bispectrum. In this case measurements of the small scales add minimal additional constraining power. In contrast, the information on \emph{equilateral} and \emph{orthogonal} PNG is primarily accessible through the bispectrum. For these shapes, small scale measurements increase the constraining power of the halo bispectrum by up to $\times4$, though the addition of scales beyond $k\approx 0.3 \mathrm{h/Mpc}$ improves constraints largely through reducing degeneracies between PNG and the other parameters. These degeneracies are even more powerfully mitigated through combining power spectrum and bispectrum measurements. However even with combined measurements and small scale information, \emph{equilateral} non-Gaussianity remains highly degenerate with $\sigma_8$ and our bias model.
\end{abstract}

\pacs{Valid PACS appear here}% PACS, the Physics and Astronomy
%\maketitle

\section{Introduction}
\label{sec:intro}
In the coming decade, a range of deep and wide photometric and spectroscopic, e.g. Rubin Observatory, DESI, SPHEREX, Euclid and Roman \citep{LSST_2009,Euclid_2011,dore_2013,Spergel_2015,DESI_2016}, galaxy surveys will produce catalogs detailing the properties of tens of billions of galaxies. Through analyzing the spatial distribution of these galaxies, and how this distribution evolves over time, we will advance our understanding in many areas of physics, from constraining the properties of neutrinos to characterizing dark energy to clarifying a range of tensions hinted at in current cosmological observations \citep[see e.g.][for overviews]{neutrino_2019,darkenergy_2019,Relics_2019,DiValentino_2021}.

A particularly exciting prospect is what can be learnt about the primordial universe. Information on the primordial physics can be extracted by studying the statistical properties of the primordial perturbations and an especially powerful avenue is through primordial non-Gaussianity (PNG), which characterizes how the distribution of the primordial perturbations deviates from Gaussianity. Through constraining PNG we help characterize the physics of the early universe, for example by learning about the field content, strength of interactions and more \citep{Maldacena_2003,Creminelli_2004,Alishahiha_2004,Senatore_2010} - see e.g. \citet{Chen_2007,Meerburg_2019} for overviews. This paper is the third in a series of papers \citep{Coulton_2022a,Jung_2022}  focused on examining what we can learn about PNG from measurements of the large scale structure of the Universe (LSS) from non-linear scales. This work focuses on an examination on assessing how much information is gained by combining binned bispectrum measurements of the halo field with power spectrum measurements. In our companion paper, Jung et al (in prep), we complement this wokr with a modal bispectrum analysis at higher redshift, demonstrating our conclusions apply more broadly, and we develop a quasi-maximum likelihood estimator, demonstrating near optimal and unbiased compression of the bispectrum. These pair of papers bring our analysis one step closer to modelling observations.

As cosmic microwave background (CMB) measurements \citep{Komatsu_2001,hinshaw2012,planck2016-l09} have limited the primordial universe to be weakly non-Gaussian, it is instructive to search for primordial bispectra, the harmonic equivalent of three-point functions. The bispectrum is the lowest order deviation from a Gaussian distribution \citep{Taylor_2000,Babich_2005} and for many classes of theoretical models the bispectrum will be the most sensitive probe of PNG. Searches for primordial bispectra typically focus on searching for theoretically motivated bispectra, known as templates or shapes, as this allows for the most statistical powerful and physically informative inferences \citep{Komatsu_2005,Babich_2005,Yadav_2007}. In this work, we focus on four shapes of the primordial bispectrum: \emph{local}, \emph{equilateral}, \emph{orthogonal-LSS} and \emph{orthogonal-CMB} \citep[the last two are two approximations for a bispectrum that arises in the effective field theory of inflation][]{Cheung_2008,Cheung_2008b,Senatore_2010} . We refer the reader to \citet{Achucarro_2022} for a recent overview of the theoretical and observational status of PNG. 

Biased tracers of the LSS, such as galaxies or halos, are potentially very powerful probes of PNG as they exhibit a novel signature of PNG, a large-scale scale dependence in the bias, that is neither seen in the matter field nor generated by many other physical processes \footnote{Neutrinos also lead to a scale dependent bias e.g. \citet{Chiang_2018},however this can be distinguished due to the different $k$ scaling}. Specifically, \citet{Dalal_2008} found that in the presence of \emph{local} PNG the bias of halos on large scales, which otherwise is expected to be approximately constant, scales as $\sim 1/k^2$. These results were subsequently expanded to show that other types of PNG can also induce scale dependent biases, with the degree of the power law behaviour determined by the squeezed limit of the primordial bispectrum, $B(k_1,k,k)$ with $k>>k_1$ \citep{Matarrese_2008,Schmidt_2010}. This feature provides new observational avenues for constraining PNG as measurements of the power spectrum can now be used to constrain PNG. The power of this can be dramatically enhanced through the use of sample variance cancellation techniques, which in principle allow types of PNG with scale dependent biases to be measured to a level limited only by the shot-noise level \citep{Seljak_2009,Castorina_2018,Chan_2019}. 

Given the prominent $\sim 1/k^2$ feature of \emph{local} non-Gaussianity, several groups have attempted to measure this using measurements of quasars and from spectroscopic galaxy surveys, such as BOSS \citep{Slosar_2008,Ross_2013,Leistedt_2014,Mueller_2021,Cabass_2022b,DAmico_2022}\footnote{Note that other complementary methods that use the mass function \citep{Lucchin_1988,Matarrese_2000,Scoccimarro_2004,Shandera_2013,Sabti_2021} have also been used to constrain \emph{local} PNG}. Whilst the best of these constraints \citep{Cabass_2022b,DAmico_2022} is not yet competitive with CMB based constraints \citep{planck2016-l09}, constraints from a broad range of surveys are forecasts to provide significantly tighter constraints on \emph{local} PNG that should surpass CMB based constraints \citep[see Fig. 5 of Ref. ][for a compendium of forecasts]{Achucarro_2022}. For \emph{local} PNG, future possible detections will likely not be limited by sample variance but rather the shot-noise of the samples and the difficulty of understanding large scale systematic effects \citep{Ross_2011,Pullen_2013,Rezaie_2021}. Note however that, as has been highlighted in \citet{Barreira_2020,Barreira_2022,Barreira_2022b}, converting any detection of PNG to a specific amplitude of the primordial bispectrum may be challenging without advances in our understanding of galaxy formation. The importance of this in ruling out single field inflationary models depends on the  exact value predicted by single field inflation;  whether it is zero or slow-roll suppressed  is continues to be subject to a theoretical discussion  \citep{Pajer_2013,Matarrese_2021}.

On the other hand, measuring other interesting shapes of PNG, such as \emph{equilateral} or \emph{orthogonal} which do not generate strong scale dependent biases, is generally significantly more challenging for LSS measurements than equivalent CMB measurements. These constraints are driven by measurements of the LSS bispectrum and are highly challenging as the late time non-linear evolution generates similar bispectra to the primordial ones \citep[see e.g.][for details of the measurements of non-linear bispectrum with BOSS ]{Gil-Marin_2015a,Gil-Marin_2015b,Gil-Marin_2016,Slepian_2017a,Slepian_2017b} . Recently there have been several new developments, first \citet{Cabass_2022a,DAmico_2022} have used effective field theory methods to robustly produce constraints on these types of PNG from measurements of the BOSS bispectrum. However just applying these methods to larger volumes will likely be insufficient to vastly improve upon CMB constraints and, as is discussed in \citet{Philcox_2022}, pushing these theoretical models deeper into the non-linear regime is highly complex. A second avenue, presented in \citet{Baumann_2021} notes that, given the inherently different physics of the processes generating these effects, it may be possible to use field level analyses or higher order correlation functions to overcome these degeneracies. 

Our work seeks to address two questions: firstly we aim to quantify how much information on \emph{local}, \emph{equilateral} and \emph{orthogonal} non-Gaussianity can be gained by extending redshift-space measurements of the halo power spectrum and bispectrum to small, non-linear scales. To do this we use the suite of \textsc{quijote-PNG} simulations presented in \citet{Coulton_2022a}. By using simulations, instead of perturbative methods, we have an accurate (dark-matter only) theoretical model into the non-linear regime and so can estimate the information content available in smaller scales. Whilst our approach imposes several assumptions, for example on the form of the halo bias, it allows us to assess an upper limit on the information content to guide future works. Second, we are interested in assessing whether there is more information about \emph{local} PNG beyond that accessible to power spectrum measurements of the scale dependent bias. \citet{dePutter_2018} first investigated this question analytically finding that for samples with moderate or low sample number densities, there is minimal extra information beyond scale dependent bias. Our simulations allow us to test this result and relax the perturbative modeling assumptions. This work builds on the first papers in this series \citep{Coulton_2022a,Jung_2022}, where we generated a large suite of simulations containing PNG, developed and validated optimal compressed estimators and used a range of methods to assess the information accessible with power spectrum and bispectrum measurements of the matter field. Our work is part of a large community interest in using small scales to constrain PNG \citep[e.g.]{Uhlemann_2018,Friedrich_2020,Cole_2020,Biagetti_2022,Giri_2022,Andrews_2022}

This paper is structured as follows: in Section \ref{sec:scale_dep_bias} we outline the primordial bispectra considered in this work and briefly review  scale dependent bias effect, and in Section \ref{sec:simulations} we describe the simulations used in this work. Then in Section \ref{sec:statistics} we overview the implementations of our power spectrum and bispectrum measurements, presenting the resulting measurements subsequently in Section \ref{sec:pk_bk_measurements}.
Next, in Section \ref{sec:FisherMethods}, we describe the details of our Fisher forecast setup, including two methods we use to enhance the robustness of our results. Finally we explore the information content of our measurements in Section \ref{sec:information_content} before presenting our conclusions in Section \ref{sec:Conclusions}. In the Appendix \ref{app:derivConvergence} we examine the stability of the derivatives estimates and, finding the standard methods to be unconverged, implement two methods that mitigate this issue thereby allowing robust constraints. In Appendix  \ref{app:covMat_convergences}  we examine the stability of our results with respect to variations in the number of simulations used to estimate the covariance matrix. Finally in Appendix \ref{app:mnu_constraints} we use the technqiues developed in Appendix \ref{app:derivConvergence} to reexamine past constraints on the sum of the masses of neutrinos.

\section{Primordial non-Gaussianity and Scale Dependent Bias} \label{sec:scale_dep_bias}
The focus of this paper is primordial bispectra defined as
\begin{align}
 &  \langle \Phi(\mathbf{k}_1)\Phi(\mathbf{k}_2)\Phi(\mathbf{k}_3)\rangle \nonumber \\ & = (2\pi)^3\delta^{(3)}(\mathbf{k}_1+\mathbf{k}_2+\mathbf{k}_3) B^{\Phi}({k}_1,{k}_2,{k}_3),
\end{align}
where $\Phi(\mathbf{k})$ is the primordial potential. In this work we consider the same four types of primordial bispectrum simulated and discussed in \citet{Coulton_2022a}. In the following, we define the relevant primordial bispectra and refer the reader to \citet{Chen_2010,Meerburg_2019,Achucarro_2022} reviews on the physics of primordial non-Gaussianity and to \citet{Coulton_2022a} for more details on the bispectra considered here.

We consider four shapes: the \emph{local} shape, with bispectrum
\begin{align}
    B^{\mathrm{Local}}_{\Phi}(k_1,k_2,k_3) = & 2 f_{\mathrm{NL}}^{\mathrm{Local}} P_\Phi(k_1)P_\Phi(k_2)+  \text{ 2 perm.},
\end{align}
the \emph{equilateral} shape, with bispectrum
\begin{align}
  &   B^{\mathrm{Equil.}}_{\Phi}(k_1,k_2,k_3) = 6 f_{\mathrm{NL}}^{\mathrm{Equil.}}\Big[- P_\Phi(k_1)P_\Phi(k_2)+\text{ 2 perm.} \nonumber \\ &  
  -2 \left( P_\Phi(k_1)P_\Phi(k_2)P_\Phi(k_3) \right)^{\frac{2}{3}} +  P_\Phi(k_1)^{\frac{1}{3}}P_\Phi(k_2)^{\frac{2}{3}}P_\Phi(k_3)  \nonumber \\ & + \text{5 perm.}\Big],
\end{align} and two approximations to the \emph{orthogonal} bispectrum - the \emph{orthogonal-LSS} with bispectrum
\begin{align} \label{eq:bis_or_lss}
   & B^{\mathrm{Orth-LSS}}_\Phi(k_1,k_2,k_3) = \nonumber \\ & 6 f_{\mathrm{NL}}^{\mathrm{Orth-LSS}}
        \left(P_\Phi(k_1)P_\Phi(k_2)P_\Phi(k_3)\right)^{\frac{2}{3}}\Bigg[ \nonumber \\ &  -\left(1+\frac{9p}{27}\right) \frac{k_3^2}{k_1k_2} + \textrm{2 perms} +\left(1+\frac{15p}{27}\right)  \frac{k_1}{k_3} \nonumber \\ &   + \textrm{5 perms}  -\left(2+\frac{60p}{27}\right)  \nonumber \\ & +\frac{p}{27}\frac{k_1^4}{k_2^2k_3^2} + \textrm{2 perms}  -\frac{20p}{27}\frac{k_1k_2}{k_3^2}+ \textrm{2 perms} \nonumber \\ & -\frac{6p}{27}\frac{k_1^3}{k_2k_3^2} + \textrm{5 perms}+\frac{15p}{27}\frac{k_1^2}{k_3^2} + \textrm{5 perms}
    \Bigg],
\end{align}
and the \emph{orthogonal-CMB}, with bispectrum
\begin{align}
    B^{\mathrm{Orth-CMB}}_\Phi&(k_1,k_2,k_3) = 6 f_{\mathrm{NL}}^{\mathrm{Orth-CMB}}\Big[-3 P_\Phi(k_1)P_\Phi(k_2) \nonumber \\ &  
   +\text{ 2 perm.}  -8 \left( P_\Phi(k_1)P_\Phi(k_2)P_\Phi(k_3) \right)^{\frac{2}{3}} +  \nonumber \\ & 3P_\Phi(k_1)^{\frac{1}{3}}P_\Phi(k_2)^{\frac{2}{3}}P_\Phi(k_3)  + \text{5 perm.}\Big],
\end{align} 
where
\begin{align}
    p=\frac{27}{-21+\frac{743}{7(20\pi^2-193)}} \,
\end{align}
where $P_\Phi(k_1)$ is the primordial power spectrum and $f_{\mathrm{NL}}^{X}$ is the amplitude of non-Gaussianity for shape $X$. The \emph{orthogonal-LSS} bispectrum is the more physically motivated shape, see \citet{Senatore_2010} for a detailed discussion, and so we primarily discuss this orthogonal shape. However, the \emph{orthogonal-CMB} shape exhibits interesting novel features that we also discuss.

Here we briefly overview of the scale dependent bias effect and refer the reader to \citet{Slosar_2008,Matarrese_2008,Desjacques_2011,Schmidt_2010} for more detailed descriptions.
The halo density contrast, $\delta_h(\mathbf{k} | M,z)$, is defined as
\begin{align}
 \delta_h(\mathbf{x}|M,z) = \frac{n_h(\mathbf{x}|M,z )}{\bar{n}_h(M,z)}-1  
\end{align}
where $n_h(\mathbf{x}|M,z)$ is the halo counts of halos with mass $M$ at redshift $z$ and $\bar{n}_h(M,z)$ is its mean. In the absence of PNG and on the largest scales, the halo overdensity is related to the matter overdensity, $\delta_m(\mathbf{k})$ by
\begin{align}
 \delta_h(\mathbf{k}|M,z) = b_1(M,z) \delta_m(\mathbf{k},z)+\epsilon(z) 
\end{align}
where $b_1 = \frac{1}{\bar{n}_h} \frac{\partial n_h}{\partial \delta_m }$ characterizes the response of the halo counts to the large scale matter fluctuation and $\epsilon(z)$ is a shot noise term from the discrete nature of the halos. 

Primordial non-Gaussianity can introduce a coupling between large and small scale modes. The most well known example is \emph{local} PNG in which small scale modes, $k_s$, of the primordial potential, are modulated by long wavelength modes, $k_l$ as \citep[e.g.][]{Komatsu_2001}
\begin{align}
\Phi(\mathbf{k}_s) \approx \Phi(\mathbf{k}_s)\Big(1+ f_\mathrm{NL}^\mathrm{Local} \Phi(\mathbf{k}_l) \Big).
\end{align} 
This coupling means that the halo overdensity is no longer just a function of the large scale matter overdensity, but also the primordial potential at that point. Hence
\begin{align}\label{eq:haloMatterBias}
& \delta_h(\mathbf{k}|M,z)  \nonumber \\& =  b_1(M,z) \delta_m(\mathbf{k},z)+ {b_\phi(M,k,z) \alpha(k,z)} \Phi(\mathbf{k})+n(\mathbf{k},z)
\nonumber \\ &= \Big[b_1(M,z)+b_\phi(M,k,z)\Big] \delta_m(\mathbf{k},z)+\epsilon(z),
\end{align}
having used in the second line the relation between the late time matter overdensity and the primordial potential, $\delta_m(k,z) = \alpha(k,z)\Phi(k)$, where $\alpha(k,z)$ is the matter transfer function. (Note the transfer function in the first line is to simplify the definition of $b_\phi$). Unlike $b_1$, the new bias term, $b_\phi$, depends on scale. 

For halos we can use the peak-background split method, assuming a universal mass function \citep{Press_1974,Sheth_2001}, to compute the relation between $b_\Phi$, $b_1$ and the primordial bispectrum $B_\Phi(k_1,k_2,k_3)$. In \citet{Schmidt_2010} this was found to be
\begin{align}\label{eq:bphib1}
b_\Phi(M,z) = \frac{2 f_\mathrm{NL} \delta_c}{\alpha(k,z)}\Big( b_1-1 \Big) \frac{\sigma^2_W(k)}{\sigma^2_M}
\end{align}
where $\delta_c\approx 1.686$ is the spherical collapse threshold, $\sigma^2_M $ is the variance of the top hat smooth matter field
\begin{align}
\sigma^2_M = \int \frac{\mathrm{d}^3k'}{(2\pi)^3} W^2(k')P_\delta(k'),
\end{align}
and $\sigma^2_W$ is the smoothed variance in the presence of primordial non-Gaussianity 
\begin{align}
&\sigma^2_W = \int \frac{\mathrm{d}^3k'}{(2\pi)^3} \frac{W^2(k')P_\delta(k')B_\Phi(\mathbf{k},\mathbf{k'},\mathbf{k'}-\mathbf{k})}{2f_\mathrm{NL} (P_\Phi(k)P_\Phi(k')+2\,\mathrm{ perms.})} .
\end{align}
In these last equations, $W_M(k)$ is the Fourier transform of the top hat, the radius of which is set by the halo mass such that $R^3=\frac{3\bar{\rho}}{4\pi} M$ and $P_\delta(k)$ is the matter power spectrum.

For the case of \emph{local} PNG $\sigma^2_W(k)\sim$ constant and, given that on large scales $\alpha(k,z)\sim k^2$, this means that $b_\Phi \propto 1/k^2$. For \emph{orthogonal-CMB} $\sigma^2_W(k)\sim k$ which leads to $b_\Phi \propto 1/k$, whilst for \emph{equilateral} and \emph{orthogonal-LSS} we have $\sigma^2_W(k)\sim k^2$ and so $b_\Phi \propto 1$. Note that as the $f_\mathrm{NL}^\mathrm{local}$ correction to the power spectrum, on currently observable scales, is constrained to be small the dominant $f_\mathrm{NL}^\mathrm{local}$ correction to the halo power spectrum is the $b_\Phi$ (i.e. the $1/k^2$ term) and not the $b_\Phi^2$ ($1/k^4$) term. In fact, the $b^2_\Phi$ term only becomes important once a detection of $f_\mathrm{NL}^\mathrm{local}$ is made.%

A particularly powerful technique to boost the constraints using the scale dependent bias effect is sample variance cancellation \citet{Seljak_2009}. The intuition underpinning this method is as follows: measurements of scale dependent bias are most powerful on the largest scales, however on these scales only a very small number of modes are measurable. This means that the sample variance on such measurements is very large and limits the constraining power on PNG. However, one can utilize the fact that the bias coefficient relating the halo/galaxy field, $\delta_h$, to the underlying matter field, $\delta_m$, is deterministic: i.e. mode-by-mode $\delta_h(\mathbf{k})= b(k)\delta_m(\mathbf{k}) +n(\mathbf{k})$ (as in Eq.\ref{eq:haloMatterBias} ). Thus if we can access the information contained in the ratio of halo and matter fields, we can perform a measurement in which the stochasticity from the matter field cancels - this is powerful as this means the sample variance terms cancel, leaving only the shot-noise terms $n(\mathbf{k})$. In the idealized zero shot-noise case, this means that the bias coefficient could be measured arbitrarily well and thus PNG could be detected with arbitrarily high significance. In practice this technique will likely be performed using multiple tracers with different biases \citep[e.g][]{Seljak_2009} as the 3D matter field is not directly observable. However a range of works have proposed using CMB lensing\citep{Schmittfull_2018}, kinetic Sunyaev Zel'dovich measurements\citep{munchmeyer_2018}, and other probes to perform sample variance cancellation with proxies for the 3D matter field.

There is one complication with using measurements of scale dependent bias to constrain PNG: the analytic form for the scale dependent bias coefficient, $b_\phi$ in Eq. \ref{eq:bphib1}, is only valid for biased tracers with universal mass functions. This is not true for observables such as quasars or galaxies, as has been explicitly explored in e.g. \citet{Slosar_2008,Reid_2010,Barreira_2020,Barreira_2022b}. A conservative approach to this issue would be to marginalize over this parameter, as is discussed in \citet{Cabass_2022b,Barreira_2022b} , however the prior is very important and is currently poorly understood. Note this marginalization is necessary as breaking this degeneracy from the data itself is very challenging, as is discussed in \citet[e.g.][]{Barreira_2022b}. In applying a simulation-based approach  in a self-consistent way, we implicitly assume that we know the bias relation, Eq. \ref{eq:bphib1}, perfectly. Our results should therefore be understood as theoretical limits. The multiplicative uncertainty in the bias relation for realistic tracers can be folded in after the fact. We discuss this further in the conclusions but leave this extension to follow-up studies.

\section{Simulations}\label{sec:simulations}
In this work we use the \textsc{quijote-png} N-body simulations presented in \citet{Villaescusa-Navarro_2020,Coulton_2022a} and refer the reader to these papers for a detailed description. The simulations were run with \textsc{GADGET-III}  \citep{Gadget}  and the halos were identified using the Friends-of-friends algorithm \citep{FoF}. In our analysis we consider a single population of halos that is comprised of all halos with $M>3.2 \times 10^{13}~h^{-1}M_\odot$, which corresponds to a mean tracer density of $\bar{n}(z=0)=1.55\times 10^{-4} \mathrm{h}^3/\mathrm{Mpc}^{3}$.  We use the 15,000 simulations run at the fiducial cosmology, with the parameters summarized in Table \ref{tab:fiducial_params}, and the simulations designed for estimating derivatives, summarized in Table \ref{tab:derivative_params}  with each parameter perturbed above and below its fiducial value. There are 500 simulations, per parameter, for the positive perturbation and 500 for the negative perturbation.  
Note that in this work we restrict our analysis to the parameters in Table  \ref{tab:derivative_params}, rather than including all the parameters varied in the \textsc{quijote} suite - which include $w$, $\Omega_b$, $\sum m_\nu$, due to the convergence challenges discussed in Section \ref{sec:FisherMethods}. This does not significantly limit our results as the other parameters are, with the exception of $\Omega_b$, extensions to $\Lambda$CDM that would likely be constrained separately, whilst $\Omega_b$ would be far better measured by the CMB \citep{planck2016-l06}.

\begin{table*}
    \centering
    \begin{tabular}{c c c c c c c c c  c c c c c}
 $\Omega_m$ & $\Omega_\Lambda$  & $ \Omega_b $ & $ \sigma_8 $  & $h$ & $n_s$ &
 Box size &$ N_{\mathrm{particles}}^{\frac{1}{3}}$ & ICs & Mass resolution  \\ 
  & &  &  &   & & (Mpc/h) &  &   &  (M$_\odot$/h) \\
    \hline \hline 
 0.3175 & 0.6825 & 0.049 & 0.834 & 0.6711 & 0.9624 &
 1000 & 512 & 2LPTPNG & $6.56\times10^{11}$  \\
    \end{tabular}
    \caption{The key properties of the 15,000 simulations used to compute our covariance matrices. Note these simulations have Gaussian primordial initial conditions - i.e. $f^X_\mathrm{NL}=0$ for all shapes, $X$.  See \cite{Villaescusa-Navarro_2020} for more details on the simulations. \label{tab:fiducial_params} }
\end{table*}
\begin{table}
    \centering
    \begin{tabular}{c| c | c}
    Parameter & Lower Value & Upper Value \\ \hline\hline
    $\Omega_{m,0}$ & 0.3075 & 3275 \\
    $h$ & 0.6511 & 0.6911 \\
    $n_s$& 0.9424 & 0.9824 \\
    $\sigma_8$ & 0.819 & 0.849 \\
    $f_\mathrm{NL}^\mathrm{Local}$ & -100  & +100 \\
    $f_\mathrm{NL}^\mathrm{Equil}$ & -100  & +100 \\
    $f_\mathrm{NL}^\mathrm{Orth-LSS}$ & -100  & +100 \\
    $f_\mathrm{NL}^\mathrm{Orth_CMB}$ & -100  & +100 \\
    $\delta_b$ & -0.035  & 0.035 \\
    $M_\mathrm{min} (M_\odot/h)$& $3.1 \times 10^{13}$ &$3.3 \times 10^{13} $ \\
    \end{tabular}
\caption{
The parameter values of the simulations used to compute the numerical derivatives. For each desired derivative, we use 500 simulations with the parameter perturbed above and 500 perturbed below its fiducial value. Note $\delta_b$ is used to compute the super sample covariance and $M_\mathrm{min}$ is used to compute the derivative with respect to the halo bias. \label{tab:derivative_params} }
\end{table}

The statistical probes, described in Section \ref{sec:statistics}, are all Fourier space statistics and so to measure these we need to preprocess the particles data and halo catalogs. First we assign them to a grid with $512^3$ voxels using the 'cloud-in-cell' mass-assignment scheme \citep{Hockney_1981}. To compute the overdensity field ($\delta_m$ and $\delta_h$ for the matter and halos respectively) we subtract the mean value  from the field and then normalize by the mean. Note this mean is computed for each realization, an important subtly for the halo field. The Nyquist frequency of the grid is $1.61 \mathrm{h/Mpc}$, which is much larger than the maximum scale used in our analysis, $k=0.5  \mathrm{h/Mpc}$, and thus dramatically reduces the effects of aliasing. See e.g. \citet{Sefusatti_2016} or Appendix A of \citet{Foreman_2020} for a discussion of aliasing effects on the bispectrum. 

For the halo field we move from real space to redshift-space by displacing the halo positions using the line-of-sight component of the halo velocity, which is computed as a mass weighting of the halo's particles. We are free to choose the direction of the line of sight and we make three grids by projecting along the $x$, $y$ and $z$ axes. Whilst these three grids share the same real-space configuration - they are not perfectly correlated in redshift-space, especially on small scales, and so we can use them to further reduce the Monte Carlo uncertainty.  We then Fourier transform the grid and deconvolve the CIC window function (see \citet{Jing_2005,Sefusatti_2016}) and then measure our statistical probes.

The halo catalogs used in this work will be available at \url{https://quijote-simulations.readthedocs.io/en/latest/png.html} on acceptance of the paper.

\section{Statistical Probes} \label{sec:statistics}
In this work we combine four statistics: the matter power spectrum, the halo power spectrum monopole and quadrupole, the halo-matter cross power spectrum and the bispectrum monopole. The combined analysis of the matter, halo-matter and halo power spectrum will allow us to exploit the sample variance cancellation method described in Section \ref{sec:simulations}. In this section we describe how we compute these statistics. The matter field statistics are computed as in \citet{Coulton_2022a} and so here we described the computation of the halo and the halo-matter cross terms.
\subsection{Halo and Matter power spectrum}

The halo redshift-space power spectrum depends both on the magnitude of the wavevector and the angle to the line of sight, $\mu = \mathbf{k}\cdot \mathbf{n}$, hence $P^{hh}(k,\mu)$. To extract information available in redshift space we compute the power spectrum monopole, $P^{hh}_0(k)$ and quadurople, $P^{hh}_2(k)$, defined as
\begin{align}
P^{hh}_\ell(k) = \frac{2\ell+1}{2} \int \mathrm{d}\mu \mathcal{L}_{\ell}(\mu) P^{hh}(k,\mu),
\end{align}
where $\mathcal{L}_\ell(\mu)$ are the Legendre polynomials. 

In an analogous manner to the matter power we estimate these as
\begin{align}
\hat{P}^{hh}_{\ell}(k_i)  = \frac{2\ell+1}{2 N'_i} \sum\limits_{k_{i}-\Delta/2<|\mathbf{k}|\leq k_i+\Delta/2} \mathcal{L}_{\ell}(\mu)\delta^h(\mathbf{k}){\delta^h}^*(\mathbf{k})
\end{align}
where $N'_i$ is the estimator normalization and we use the same binning as for the matter power spectrum.  We use the public \textit{Pylians3} code \footnote{\url{https://github.com/franciscovillaescusa/Pylians3}} to compute these quantities. Additionally we subtract the shot-noise term, $1/\bar{n}$, from the halo power spectrum monopole.

Likewise, we estimate the matter-halo cross power spectrum monopole as
\begin{align}
&\hat{P}^{mh}_{\ell}(k_i)  = \frac{1}{2 N'_i} \times \nonumber \\ & \sum\limits_{k_{i}-\Delta/2<|\mathbf{k}|\leq k_i+\Delta/2} \left(\delta^m(\mathbf{k}){\delta^h}^*(\mathbf{k})+{\delta^m}^*(\mathbf{k}){\delta^h}(\mathbf{k}) \right)
\end{align}

As is discussed in Section \ref{sec:FisherMethods}, we limit the wavenumbers of the matter power spectrum and matter-halo cross power spectrum to $k<0.1\mathrm{h/Mpc}$.

\begin{figure*}
    \centering
  \includegraphics[width=0.85\textwidth]{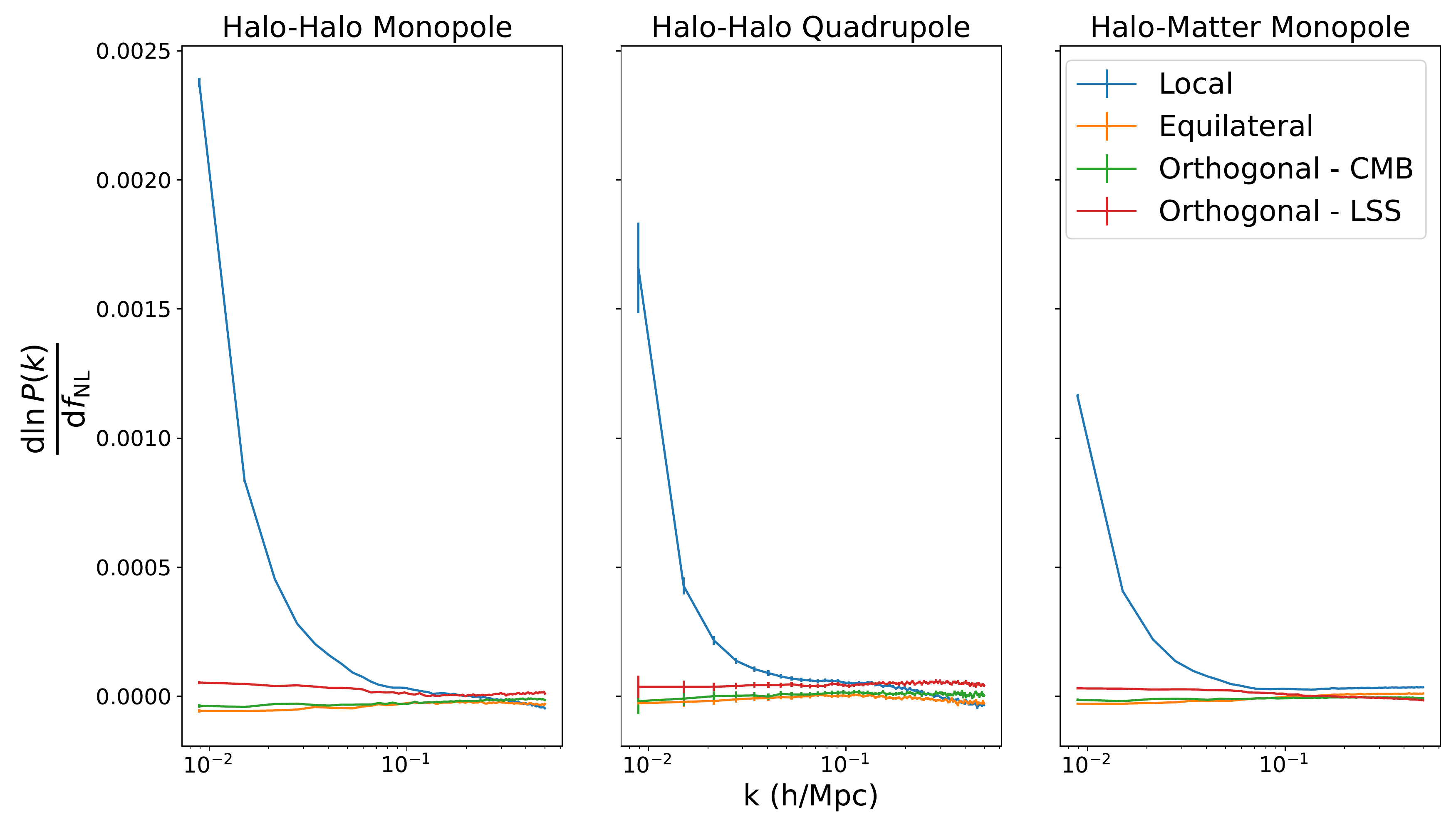}
    \caption{The derivative of the halo monopole, halo quadrupole and halo-matter monopole power spectra with respect to the amplitudes of the four types of primordial non-Gaussianity. These are computed at $z=0.0$ and the halo field contains all halos with $M>3.2 \times 10^{13} M_\odot/h$. The error bars are denote the error on the mean. The large response of \emph{local} non-Gaussainity arises due to the well-known, scale-dependent bias effect (see Section \ref{sec:scale_dep_bias}). The response of the power spectra to other types of non-Gaussianity is significantly weaker and largely featureless.  \label{fig:pk_response}}
\end{figure*}
\begin{figure}
  \includegraphics[width=0.45\textwidth]{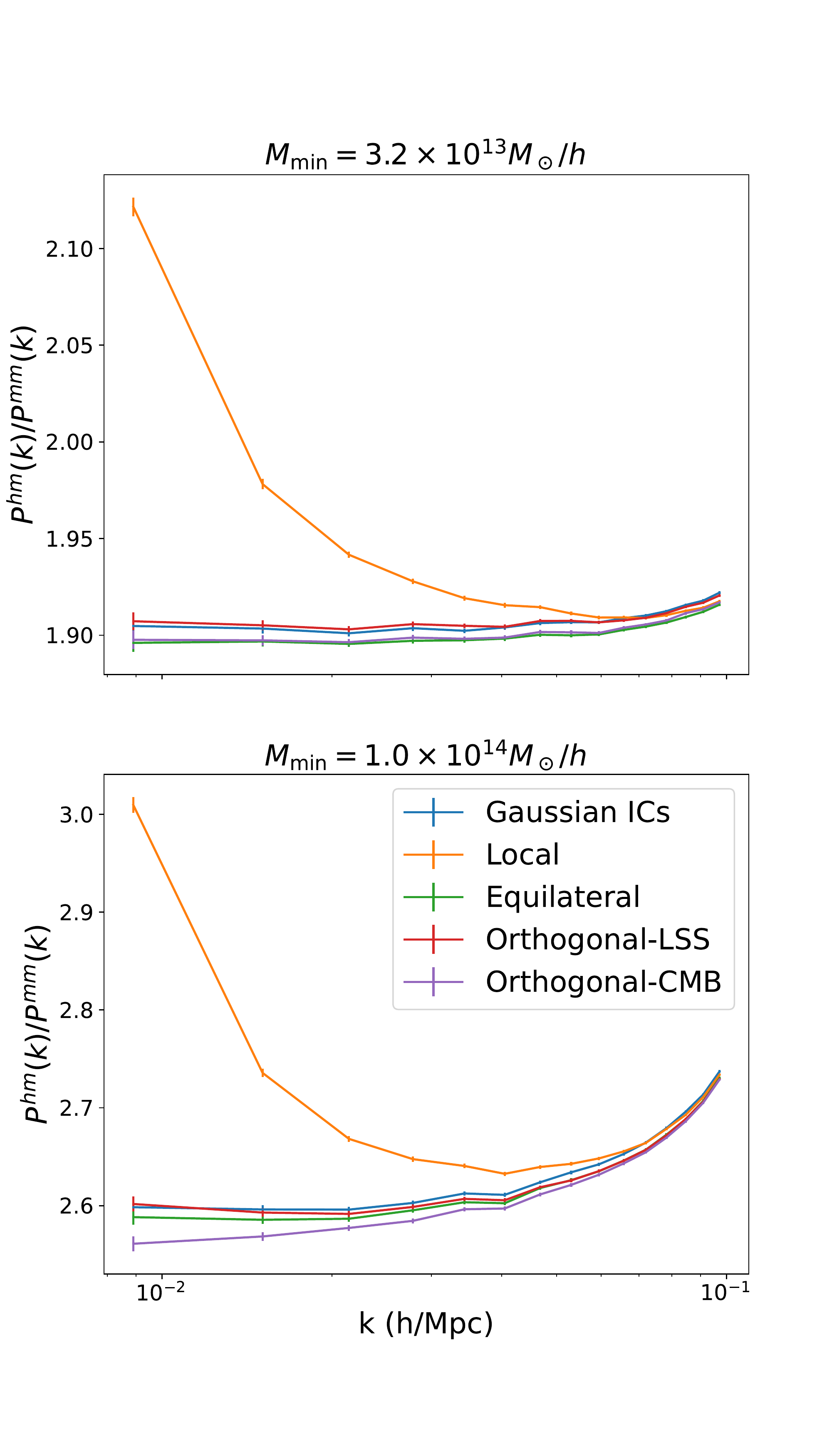}
    \caption{The average ratio of the halo-matter power spectrum to the matter-matter power spectrum for two halo mass limited samples at $z=0.0$. The scale dependent bias terms can be isolated through measurements of this ratio on large scales. These results show how amplitude of the scale dependent bias terms responses to changes in the halo sample. Interestingly we only see evidence of the expected scale dependence in the \emph{orthogonal-CMB} shape for the higher mass sample, see the text for a detailed discussion. \label{fig:ps_scaleDep} }
 \end{figure}
\begin{figure}
  \includegraphics[width=0.45\textwidth]{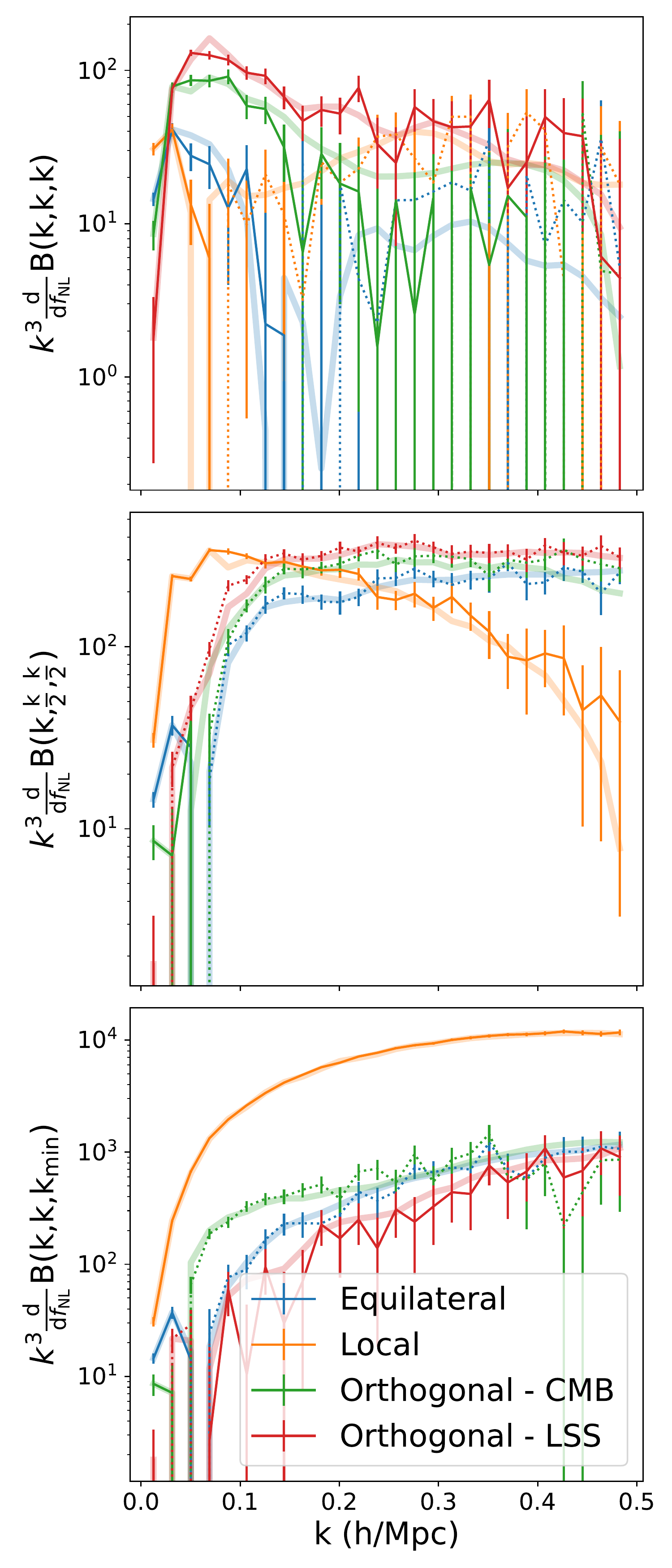}
    \caption{The response of the halo- equilateral (top), folded (middle) and squeezed (bottom) bispectrum slices to the four different types of primordial non-Gaussianity at $z=0$. The dotted lines denote regions where the bispectrum is negative and the error bars are the error on the mean of our simulations, which is equivalent to the error that would be measured with a volume of $\sim 500$ Gpc$^{3}$/h$^3$. At small scales, especially for the equilateral slice, the measurements are significantly impacted by noise, despite using 500 simulations with matched seeds. This large noise can bias our Fisher estimates and in Appendix \ref{app:derivConvergence} we demonstrate two methods to mitigate this bias. The method used in the rest of the paper mitigates the noise by fitting a smooth function, via a Gaussian process, to the measurements. The thick shaded contour shows the resulting smoothed derivatives. \label{fig:bispec_wGP} } 
 \end{figure}
\subsection{Halo bispectrum}
The halo bispectrum is defined as 
\begin{align}
\langle \delta^h(\mathbf{k}_1)& \delta^h(\mathbf{k}_2)  \delta^h(\mathbf{k}_3)  \rangle=  \nonumber \\ & (2\pi)^3 \delta^{(3)}(\mathbf{k_1}+\mathbf{k}_2+\mathbf{k}_3) B(\mathbf{k}_1,\mathbf{k}_2,\mathbf{k}_3).
\end{align}
The redshift space distortions mean that, like the halo power spectrum, the halo bispectrum also depends on the orientation of the modes relative to the line of sight. In this work we consider only the bispectrum monopole, which is defined simply averaging over all angles. Thus we use the same estimator as for the matter bispectrum.
\begin{align} 
\hat{B}^{hhh}_0(k_i,k_j,k_k) &= \frac{1}{N_{ijk} } \sum\limits_{k_{i-1}<|\mathbf{k_1}|\leq k_i,k_{j-1}<|\mathbf{k_2}|\leq k_j,k_{k-1}<|\mathbf{k_3}|\leq k_k} \nonumber \\ & \delta^h(\mathbf{k}_1)\delta^h(\mathbf{k}_2)\delta^h(\mathbf{k}_3) (2\pi)^3\delta^{(3)}(\mathbf{k}_1+\mathbf{k}_2+\mathbf{k}_3),
\end{align}

The halo bispectrum monopole contains a contribution from the discrete nature of the halo field. This term is known as the shot-noise term and in this work we remove it.  The shot-noise term is computed as 
\begin{align}
& \hat{B}^{SN}_0(k_i,k_j,k_j)=\frac{1}{N_{ijk} } \sum (2\pi)^3\delta^{(3)}(\mathbf{k}_1+\mathbf{k}_2+\mathbf{k}_3)  \nonumber \\ & \left[ \Big( P^{hh}_0(k_1) +P^{hh}_0(k_2)+P^{hh}_0(k_3) \Big) \frac{1}{\bar{n}} +\frac{1}{\bar{n}^2 } \right],
\end{align}
where the implicit limits are the same as for the bispectrum estimator and $\bar{n}$ is the mean halo number density. We use the halo power spectrum and mean halo number density as measured on that realization to compute this term. This removes the sample variance arising from the Poisson term - note however it does not remove the shot-noise contribution to the bispectrum variance.

\section{Power spectrum and bispectrum measurements}\label{sec:pk_bk_measurements}
In this section, we explore how the power spectrum and bispectrum statistics are impacted by PNG using the simulations summarized in Table \ref{tab:derivative_params}.
\subsection{Power spectrum}
In Fig. \ref{fig:pk_response}  we plot the derivative of the halo power spectrum with respect to the four shapes of primordial non-Gaussianity. For \emph{local}, the scale dependent bias effect, discussed Section \ref{sec:scale_dep_bias},  leads to an enhancement of the large scale power spectrum, as expected on these scales, as $k^{-2}$ \citep[see e.g.][ for analytical description of this behaviour]{Schmidt_2010}.  No scale dependent bias effect is observed for the other shapes of primordial non-Gaussianity. For \emph{equilateral} and \emph{orthogonal-LSS} this is expected and is consistent with previous works and theory \citep[e.g.][]{Wagner_2012,Scoccimarro_2012,Sefusatti_2012}. However the \emph{orthogonal-CMB} PNG is expected to exhibit a scale dependent bias that scales as $k^{-1}$.

This apparent inconsistency with expectations for the \emph{orthogonal-CMB} case arises as, for the halo sample used in this work, the coefficient for the scale dependent bias term, Eq. \ref{eq:bphib1}, is very small. When considering other samples the coefficient can be much larger and then the scale dependent bias can be seen.  In Fig. \ref{fig:ps_scaleDep}, we show the ratio halo-matter power spectrum to the matter power spectrum for our original sample, with a minimum mass of $M_\mathrm{min}=3.2 \times 10^{13}M_\odot /h$, and a second sample with $M_\mathrm{min}=1\times 10^{14} M_\odot /h$. On large scales this ratio isolates the linear bias term. For the second sample we see a scale dependent PNG bias contribution for both the \emph{local} and \emph{orthogonal-CMB} cases. The small correction seen for our original sample is in fact consistent with previous work: \citet{Schmidt_2010} found that for sample of halos at  $z=0.0$ and $M=1\times 10^{13} M_\odot /h$ the scale dependent bias for \emph{orthogonal-CMB} PNG is $\sim 1\%$ of the linear halo bias and more than an order of magnitude smaller than the equivalent \emph{local} shape bias. For such small values of $b_\phi$ the scale dependent bias, on the scales considered here, is swamped by the changes to $b_1$ induced by PNG.

\begin{figure}
  \includegraphics[width=0.45\textwidth]{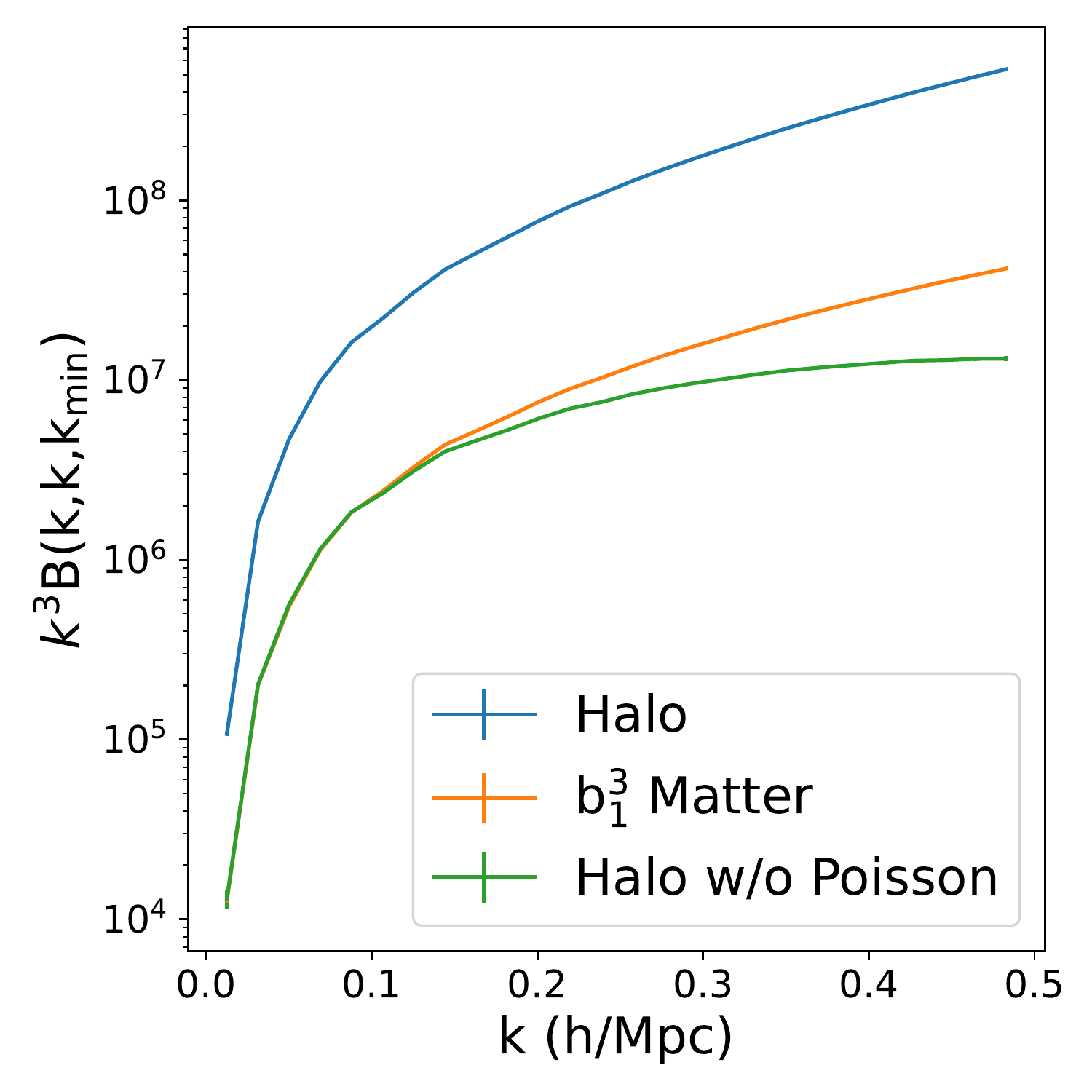}
    \caption{ The squeezed bispectrum computed at the fiducial cosmology (Table \ref{tab:fiducial_params}). We compare the halo bispectrum to the matter bispectrum rescaled by the linear bias, $b_1$, and to the halo bispectrum once we have removed the shot noise contribution. The large shot-noise contribution to the bispectrum leads to the large noise seen in our bispectrum derivatives, Fig. \ref{fig:bispec_wGP}.  \label{fig:bispec_fiducial} }
 \end{figure}

\subsection{Halo bispectrum}
In Fig. \ref{fig:bispec_wGP} we investigate the signatures of PNG in the bispectrum at $z=0.0$. On the largest scales the bispectra are all very well measured and we have tested that these are well described by a simple linear bias times the PNG contribution to the matter bispectrum, see \citet{Coulton_2022a,Jung_2022} for more details of the matter bispectrum. Moving to smaller scales we see that the halo bispectrum derivatives tends to scale as $\sim k^{-3}$; interestingly this scaling is significantly weaker than the one of the matter bispectrum, which scales as $\sim k^{-4}$. These results are qualitatively consistent with the theoretical models investigated in \citep[e.g.][]{Sefusatti_2009,Baldauf_2011} and previous simulations \citep{Sefusatti_2012,Tellarini_2016} though we push to smaller scales in this work. Secondly we notice that the bispectra are very noisy, much more so than the matter bispectra. The high levels of noise in the bispectrum derivatives, even after averaging over 500 simulations, leads to challenges when performing a Fisher forecast and we explore the details of this in Section \ref{sec:FisherMethods} and Appendix \ref{app:derivConvergence}.

To further understand the source of noise we investigate the properties of the bispectrum at the fiducial cosmology of the \textsc{quijote} . In Fig. \ref{fig:bispec_fiducial} we compare the squeezed slice of the halo bispectrum with and without shot-noise subtraction to the matter bispectrum, scaled by the linear bias. Note that the linear bias was estimated by computing the  ratio of the halo-matter cross power spectrum to the matter auto power spectrum. We see that on all scales the halo-bispectrum without the shot-noise correction is significantly larger than the other bispectra, reaching a factor of $\sim 10$ on the smallest scales. This large shot-noise acts as a source of effective noise in our analysis. When computing the derivatives shown in Fig. \ref{fig:bispec_wGP}  we use simulations with matched initial conditions: the amplitudes and phases of the initial conditions are identical and the only difference is the sign of the PNG correction term, see Section IIIA in \citet{Coulton_2022a} for more details. For the matter field, using matched simulations means the effective noise terms cancel very effectively leading to highly accurate, low noise derivative measurements (note that this is further aided as noise in the matter field is inherently small on these scales). The halo field is more complicated as the slightly different evolution of the two simulations used to estimate the derivative ($f_\mathrm{NL}=+100$ and $f_\mathrm{NL}=-100$) leads to slight differences in the number and positions of the halos. These slight differences mean that the cancellation of the shot-noise is not perfect and some of the large noise leaks in and masks the small bispectrum signals.

Note that, once the shot-noise has been subtracted, we find a good agreement between the scaled matter bispectrum and the halo bispectrum on the largest scales. This is in agreement with perturbation theory results for the largest scales \citep{Baldauf_2011}. On smaller scales we see large differences between the scaled matter and shot-noise subtracted bispectra - this is also expected as moving to smaller scales higher order bias terms become important and beyond the non-linear scale, $k\sim 0.08 \mathrm{Mpc/h}$, the perturbative description breaks down.

\section{Fisher forecasting and the challenges of convergence}\label{sec:FisherMethods}

The basic Fisher forecast methodology is very similar to \citet{Coulton_2022a} and so we outline the key steps and refer the reader to that work for more details. 

We assume that the power spectra and bispectra likelihood can be approximated by a Gaussian distribution. Thus the Fisher information is given by 
\begin{align}
    F_{IJ} = \frac{\partial\bar{O}(k_i)}{\partial\theta_I}C_{ij}^{-1}\frac{\partial\bar{O}(k_j)}{\partial\theta_J}.
\end{align}
where $O$ is the data vector (composed of all or subsets of the power spectra and bispectrum measurements), $\theta_I$ are the parameters of interest and $C$ is covariance matrix. The derivatives are estimated using central finite differences with a 2-point stencil and the covariance matrix is estimated from the simulations. 

 We focus our analysis on the four PNG bispectrum ampltidues: $f_\mathrm{NL}^\mathrm{Local}$, $f_\mathrm{NL}^\mathrm{Equil.}$, $f_\mathrm{NL}^\mathrm{Orth.-LSS}$, and $f_\mathrm{NL}^\mathrm{Orth.-CMB}$; four cosmological parameters $\Omega_m$, $n_s$, $\sigma_8$ and $h$; and one parameter controlling the halo bias, $M_\mathrm{min}$. The bias parameter, $M_\mathrm{min}$, corresponds to variations in the halo catalog mass threshold and to compute derivatives of this we generate, and process in an identical manner, catalogs with minimum halo mass of $3.1 \times 10^{13} M_\odot/h$ and $3.3 \times 10^{13} M_\odot/h$ \citep[see e.g.][for a discussion of this parameterization]{Hahn_2020}. This bias parameter is, in our setup, roughly equivalent to the standard linear bias model. Whilst the bias model presented here is overly simplistic it is still useful as first estimate of the importance of the bias terms and for estimating the total accessible information. A more thorough analysis of the impact of bias parameters, which could be implemented in our simulations by populating the halos with a HOD and investigating the constraints on the HOD parameters, is left to future work. It is expected that including more bias parameters will lead to further degradation of the constraints as is discussed in \citep{DAmico_2022b,Philcox_2022}.

We use 14,500 simulation at the \textsc{quijote} fiducial cosmology to compute the covariance (the remaining 500 are used to compute the super sample covariance terms0. Unlike for the computation of the derivatives, we only include projections along one line-of-sight in the simulations used to estimate the covariance matrix. This is to avoid biasing the covariance matrix with correlations between the correlated lines-of-sight. In Section \ref{sec:covMatContributions} we investigate the importance of the different contributions to the covariance matrix, which is composed of Gaussian contributions, non-Gaussian terms from the connected 3-,4- and 6- point functions and the super sample covariance \citep[see e.g.][]{Kayo_2013,Chan_2017,Gualdi_2018,Sugiyama_2020}.

In this work our primary focus is to investigate the information of the halo field. As a consequence, we limit the information included in the matter field to only the largest scale modes $k<0.1 \mathrm{h/Mpc}$. Through this choice we can assess the benefits of sample variance cancellation (through the matter field on the largest scales) as will be used in future surveys through multitracer analyses \citep{Slosar_2009}, with CMB lensing \citep{Schmittfull_2018} or with the kinetic Sunyaev Zel'dovich effect\citep{munchmeyer_2018}. The $k$ cut used here is larger than would be accessible through some of these methods, however this compensates for the small size of our box. Whilst using multiple tracers is a powerful method to employ sample variance cancellation; we do not use that here as, given our mass resolution, it is difficult to construct a second tracer field from our halo catalog with an observationally relevant tracer density. 

An important part of Fisher forecasts with numerically estimated covariance matrices and derivatives is that the Monte Carlo noise in these components is sufficiently small so that it does not bias the forecasts. In Appendix \ref{app:derivConvergence} we examine the stability of derivative estimates and in Appendix \ref{app:covMat_convergences} we investigate the stability of the covariance matrix. We find that our derivatives are not converged, leading to biased  forecasts when using  the standard way to compute the Fisher matrix. To mitigate this issue we implement two complementary methods to obtain stable and robust results. This is an important, but technical aspect of our forecast, and we provide a  detailed description in Appendix \ref{app:derivConvergence}. The bispectrum results presented in the remainder of the paper use the `smoothed' (Gaussian process) method to ensure robust forecasts.

\section{Constraining Power of the halo field}\label{sec:information_content}
In this section we explore the importance of degeneracies, small scale information and the different contributions to the covariance matrix when quantifying the amount of information on PNG embedded into the halo power spectrum and bispectrum.  Our analysis is performed at $z=0.0$ and uses modes up to $k_\mathrm{max}=0.5~ \mathrm{h/Mpc}$ except for quantities involving the matter field, for which we use $k_\mathrm{max}=0.1~ \mathrm{h/Mpc}$.  As discussed in Section \ref{sec:FisherMethods}, we use low $k_\mathrm{max}$ measurements of the matter field to emulate observations that use the sample variance cancellation method and assess how much information they contain. The results in this section primarily focus on the \emph{local}, \emph{equilateral} and \emph{orthogonal-LSS} shapes as these most closely approximate physically generated bispectra. 
\subsection{Parameter degeneracies}
\begin{figure*}
  \includegraphics[width=1.\textwidth]{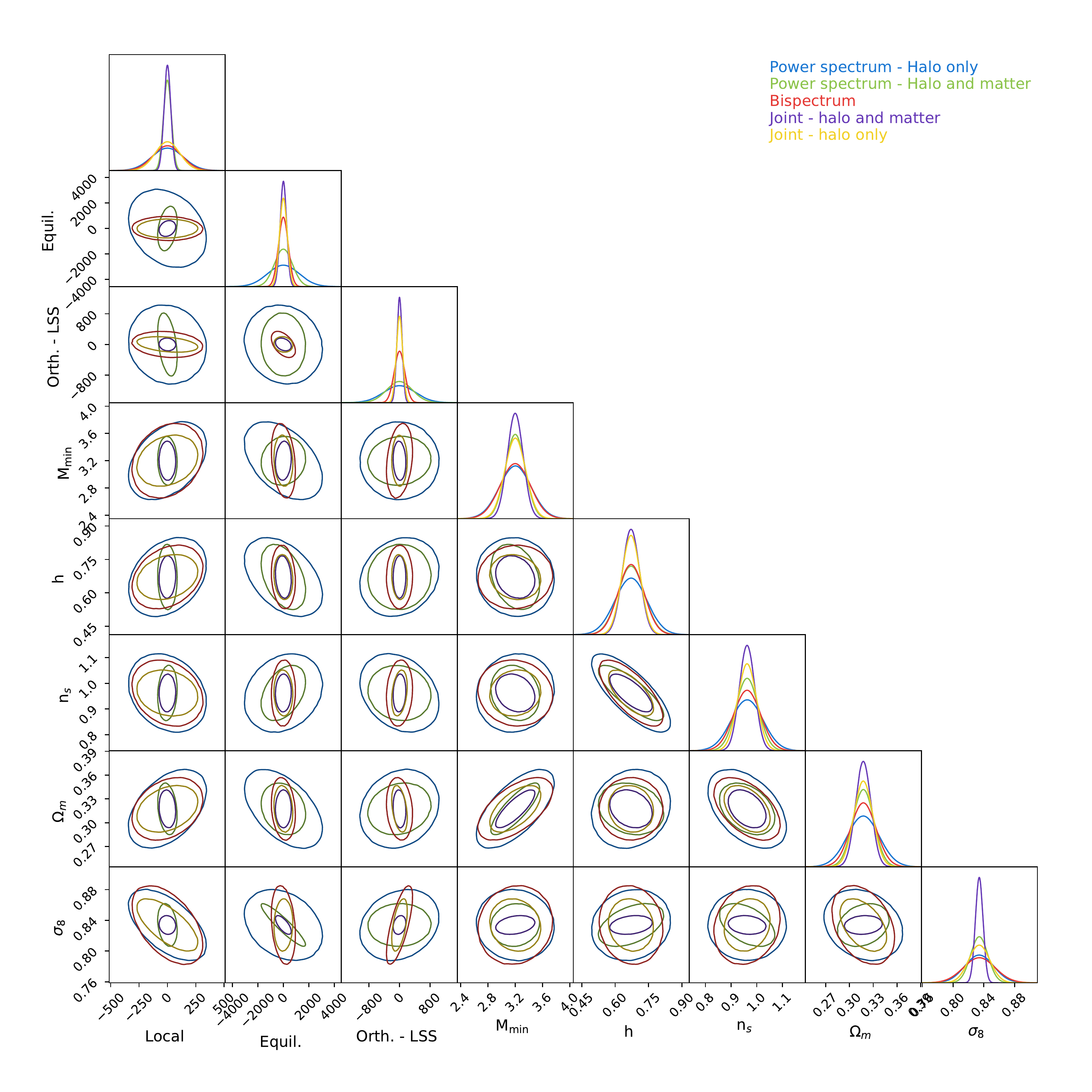}
    \caption{ A comparison of the constraining power in the halo power spectrum monopole and quadrupole, the matter power spectrum, the halo-matter power spectrum monopole and the halo bispectrum. This analysis is performed at $z=0.0$ with a maximum scale of $k_\mathrm{max} = 0.5 \mathrm{h/Mpc}$ except for the matter field, for which we only include modes up to  $k_\mathrm{max} = 0.1 \mathrm{h/Mpc}$. These results are for a 1 Gpc$^{3}$/h$^3$ volume at $z=0.0$ for tracers with a number density of $1.55\times 10^{-4} \mathrm{h}^3/\mathrm{Mpc}^{3}$. The contours show the 2$\sigma$ constraints. These results demonstrate the sensitivities and different degeneracies directions for each of the probes. \label{fig:param_const_corner} } 
 \end{figure*}
 \begin{figure*}
    \centering
  \subfloat[Halo field only ]{\label{fig:constraints_kmax_haloOnly}%
  \includegraphics[width=0.95\textwidth]{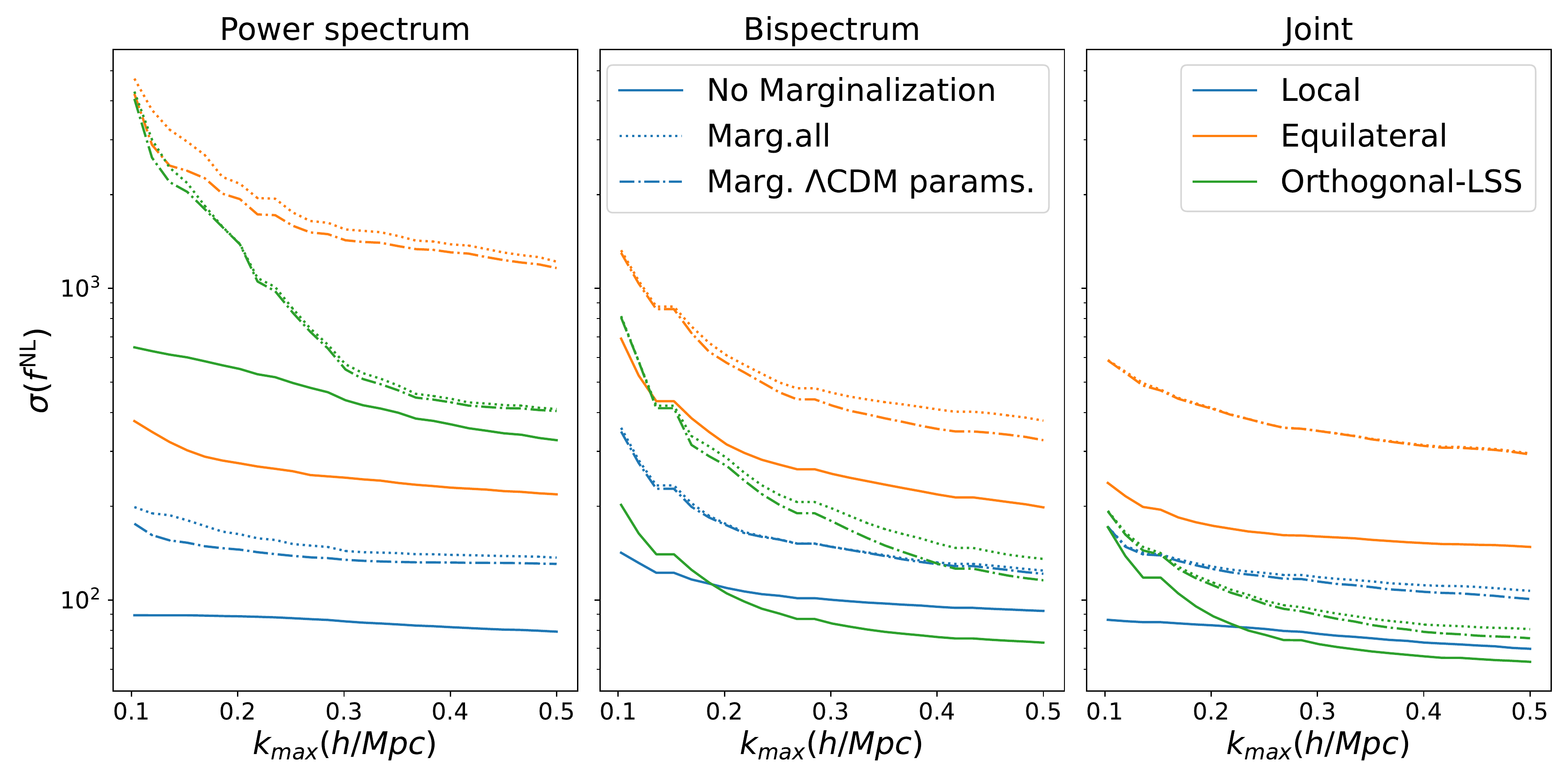}%
 }
 \\
    \subfloat[The halo field statistics combined with measurements of the large scale ($k<0.1 \mathrm{h/Mpc}$ ) matter power spectrum and halo-matter cross power spectrum.]{\label{fig:constraints_kmax}%
  \includegraphics[width=0.95\textwidth]{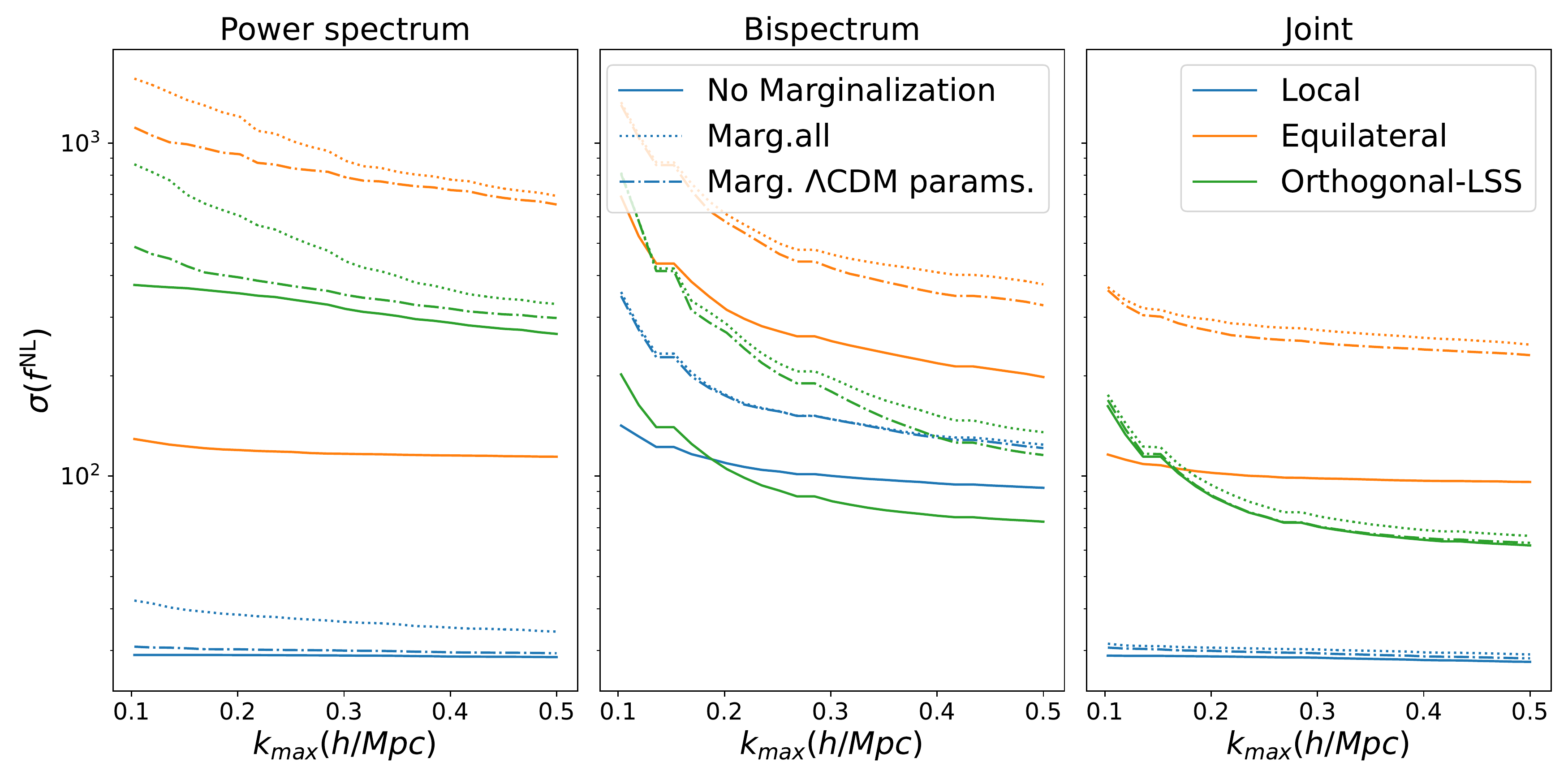}%
}  
\caption{An examination of how the constraining power for power spectrum and bispectrum measurements vary as a function of the maximum scale used in the analysis. The solid lines corresponds to the 1-$\sigma$ constraining power when no marginalization is performed, the dash-dotted includes marginalization of the cosmological and bias parameters and the dotted line corresponds to marginalization of cosmological and bias parameters, and the other PNG shapes. These are compute at $z=0.$ with the same tracer as in Fig. \ref{fig:param_const_corner}. These results demonstrate the value of including scales beyond the perturbative regime, $k\gtrsim 0.1 \mathrm{h/Mpc}$. The inclusion of the large scale matter field, Fig. \ref{fig:constraints_kmax}, to the halo field only results, Fig. \ref{fig:constraints_kmax_haloOnly}, highlights the power of sample variance cancellation methods.}
\end{figure*}
 In Fig. \ref{fig:param_const_corner} we explore the parameter constraints obtainable from the bispectrum and power spectrum measurements. First, focusing on the power spectrum constraints, we see that, compared to the information available in measurements of the matter field power spectrum discussed in \citet{Coulton_2022a,Jung_2022}, the halo power spectrum contains much more information on PNG. This is largely due to the scale dependent bias effects. The information in scale dependent bias can be more efficiently extracted with sample variance cancellation techniques, as originally proposed in \citet{Seljak_2009}. In our analysis, we attempt to incorporate this effect by using both the monopole, matter monopole and halo-matter cross power spectra. It can be seen that including these other power spectrum measurements vastly improves the constraints on PNG, particularly for \emph{local} non-Gaussianity.
 
Moving to the bispectrum constraints, we see that the constraints on \emph{local} PNG are worse for the bispectrum than for the power spectrum - highlighting the importance of scale dependent bias measurements. As expected, for the \emph{equilateral} and \emph{orthogonal} shapes the bispectrum measurements are significantly more informative than the power spectrum measurements. We see that there are strong parameter degeneracies for the bispectrum measurements of PNG - especially with the effective bias parameter $M_\mathrm{min}$ and the amplitude of clustering, $\sigma_8$. The degradation of the constraints due to marginalization can be seen more clearly in Fig. \ref{fig:constraints_kmax}, where it can be seen that for the bispectrum marginalization increases the constraints by more than $100\%$. 

Finally we see that combining the power spectrum and bispectrum measurements provides significant improvements for all the parameters, except \emph{local} PNG where the gains are more modest. These improvements come from two related effects: the power spectrum and bispectrum exhibit different degeneracies between the PNG parameters and the $\Lambda$CDM parameters and second through the complementary nature of the two probes: subsets of the parameters are better constrained by different probes and when combined this improved constraining power propagates to the other parameters due to the strong degeneracies.

\subsection{Are small scale measurements informative?}

To further understand the information content, it is useful to examine which scales contribute to the constraining power. In Fig. \ref{fig:constraints_kmax_haloOnly} we explore the constraining power in measurements of the halo field only as a function of scale. For the unmarginalized constraints, those obtained when measuring only the amplitude of one type of PNG, the information in the power spectrum changes very slowly with the inclusion of new modes whereas the bispectrum constraints improve rapidly with $k_\mathrm{max}$ up to $k_\mathrm{max} \sim 0.3~ \mathrm{h/Mpc}$ after which the gains are more modest. This shows that, like the matter field \citep{Coulton_2022a,Jung_2022}, there is minimal gains from pushing to very small scales - at least up to the maximum scale considered here. When considering marginalized constraints the situation is somewhat different: the power spectrum and bispectrum constraints both show significant improvements up to $k_\mathrm{max}  \sim 0.4~\mathrm{h/Mpc}$. This improvement arises as the inclusion of these scales reduces the degeneracies and improves the constraints on the degenerate parameters. Finally when we consider the joint analysis of the power spectrum and bispectrum, we see that the complementarity of the probes helps reduce the degradation of the constraints from marginalization. Interestingly, we find the joint marginalized constraints for \emph{local} and \emph{orthogonal-LSS} show minimal degradation from marginalization, and thus the joint case mirrors the unmarginalized case where there is minimal information in the small scale modes $k\gtrsim 0.3~ \mathrm{h/Mpc}$.

Next we repeat this analysis for the case where we include large scale ($k<0.1~\mathrm{h/Mpc}$) measurements of the matter auto and matter-halo cross power spectra. By comparing with Fig. \ref{fig:constraints_kmax_haloOnly}, we see the power of sample variance cancellation (which occurs on large scales through the inclusion of the matter field) as the constraints improve by a factor of $\sim 5$. Note that, in the unmarginalized case, sample variance cancellation improves the power spectrum constraint for \emph{equilateral}, despite the absence of a scale dependent bias feature. This occurs as equilateral non-Gaussianity leads to a small shift in the linear bias on large scales, see Fig. \ref{fig:ps_scaleDep}. As we cannot predict the linear bias coefficient this information cannot be accessed in practice. This effect is completely degenerate with halo bias so when we marginalize over the bias model the constraints dramatically degrade. The evolution with $k_\mathrm{max}$ otherwise mirrors the halo-only analysis: the unmarginalized power spectrum constraints improve weakly with the inclusion of small scale modes whilst the marginalized case shows significant improvements as the small scale modes, up to $k_\mathrm{max}\sim 0.3~ \mathrm{h/Mpc}$, are included. Likewise, the joint analysis of both probes also shows only modest improvements with $k_\mathrm{max}$. 

\begin{figure}
  \includegraphics[width=0.45\textwidth]{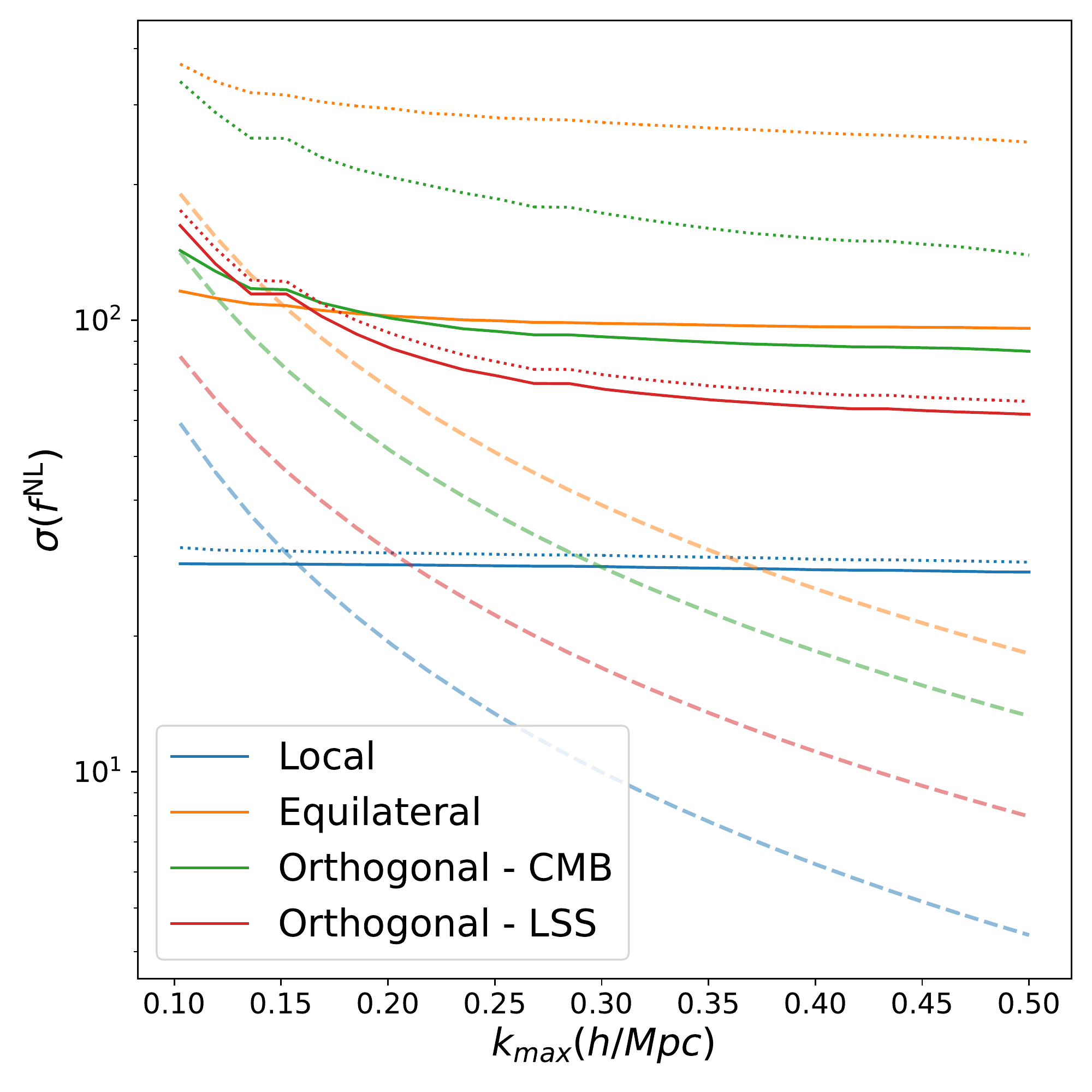}
    \caption{ A comparison of the information in the primordial fields (dashed) to that accessible with joint measurements of the matter and halo power spectra and the halo bispectrum monopole (solid lines denote the unmarginalized constraints and the dotted are the marginalized constraints).  The marginalization includes the marginalization of the $\Lambda$CDM parameters, the bias parameter and the other PNG templates. Note however that we only marginalize over one of the orthogonal templates rather than both as these are two different approximations of the same physical bispectrum. \label{fig:constraints_kmax_primordial} } 
 \end{figure}

\begin{figure}
  \includegraphics[width=0.45\textwidth]{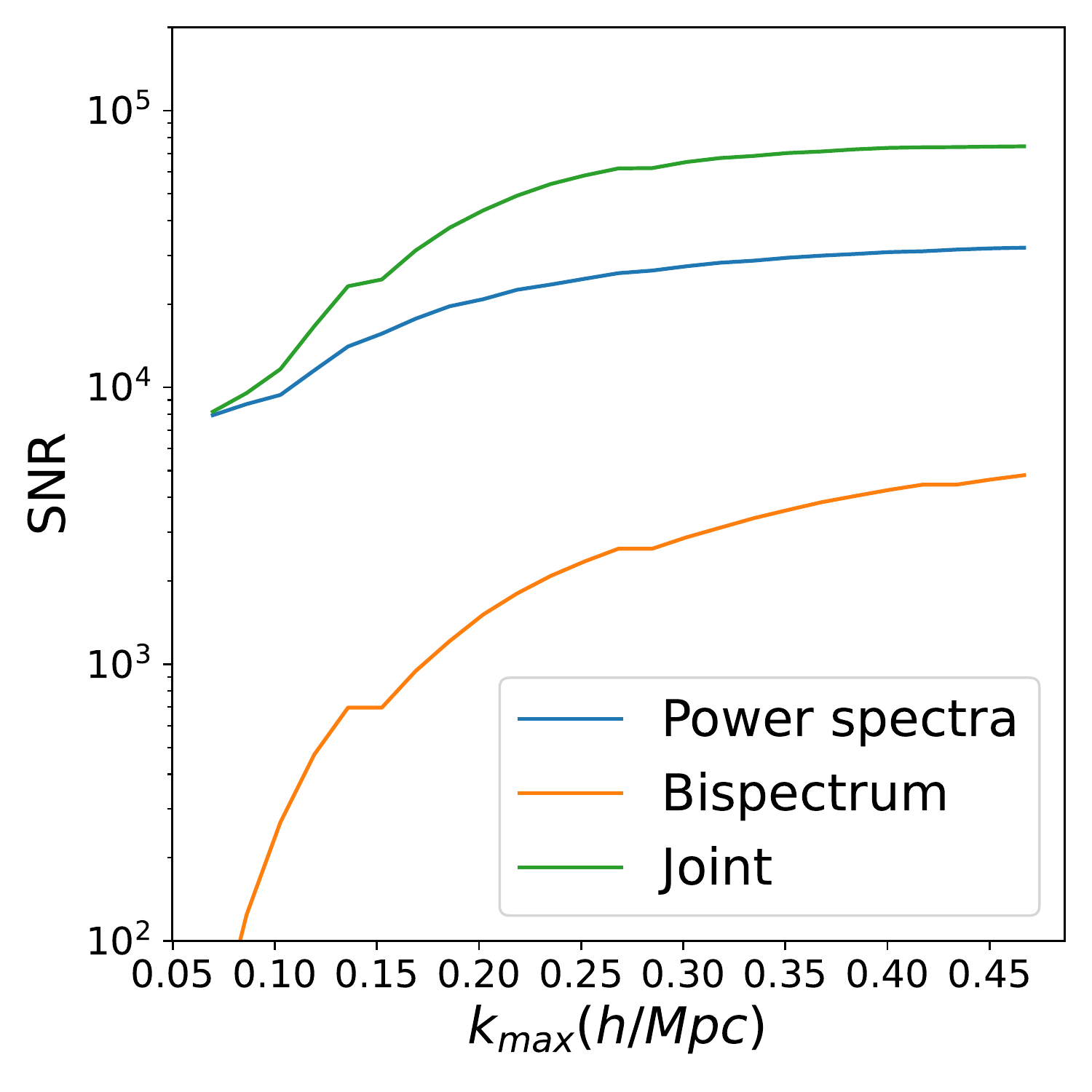}
    \caption{ The cumulative signal to noise ratio (SNR) of the combined halo and matter power spectra, the bispectrum and the combination of all the probes. At small scales the modes become increasingly correlated, meaning that the increase in SNR is significantly less than the naive scaling of $k^\frac{3}{2}$. \label{fig:snr_plot} } 
 \end{figure}

To contextualize these results, in Fig. \ref{fig:constraints_kmax_primordial} we compare the information available in the joint power spectrum and bispectrum analysis to the information available in the primordial fields. There are several interesting features: firstly the constraints on \emph{local} PNG on the largest scales are better in the late time than in the primordial universe. This result is easily explained as the information encoded in the scale dependent bias, which drives the constraints on the largest scales, actually encodes information from small scales - the formation of a halo is inherently controlled by these modes and the information of these small scale modes is transferred to the halo bias. Note that we discuss the challenging issue of marginalizing over $b_\phi$ in the conclusions, Section \ref{sec:Conclusions}. Next we can see a quantification of the saturation of the constraints: in the primordial space these improve at least as fast as $k_\mathrm{max}^{\frac{3}{2}}$  \citep{Kalaja_2021} whereas the late time improvement is drastically reduced. This saturation was also seen in the matter field \citep{Coulton_2022a} and in a similar manner arises as the SNR ratio of the probes, shown in Fig. \ref{fig:snr_plot}, improves slowly. We discuss the origin of this saturation in the next section.
 
\subsection{Sensitivity to modeling of the covariance matrix}\label{sec:covMatContributions}
The covariance matrix of our probes can be written as 
\begin{align}\label{eq:covMatContributions}
C_{ij} = C^\mathrm{Gaus.}_{ij}+C^\mathrm{non-Gaus.}_{ij}+C^\mathrm{SSC}_{ij},
\end{align}
where the three terms are: the Gaussian (or disconnected), the non-Gaussian (or connected) and super sample covariance (SSC) contributions. Whilst the contribution of the Gaussian term, see Appendix B in \citet{Coulton_2022a} for explicit expressions, is straightforward to  calculate analytically, the remaining two terms are complex and typically need to be estimated from simulations \citep[see e.g][]{Kayo_2013,Chan_2017,Gualdi_2018,Sugiyama_2020}. Given the large size of our data vector, with 2,370 elements for the joint bispectrum and halo-matter power spectrum analysis, we require at least 2,370 simulations in order for the estimated covariance matrix to be invertible. From Appendix \ref{app:covMat_convergences} we see that we in fact need at least 3,000 for our forecast constraints to be converged at the $ 10\%$ level. 

Generating thousands of simulations is computationally very expensive and hence we investigate the necessity of including the non-Gaussian and SSC terms. We compute the SSC terms using the separate universe method described in \citet{Li_2014}. We use 1,000 simulations (500 with a large scale overdensity mode and 500 with a large scale underdensity mode, see Appendix B in \citet{Coulton_2022a} for explicit formulae). This work builds on the results from many previous works, \citep{Chan_2017,Chan_2018,Barreira_2019,Biagetti_2021} which have found that the non-Gaussian and SSC terms can be the dominant contribution to the covariance matrix, by propagating the impact of these terms into the parameter constraints.

\begin{figure}
  \includegraphics[width=0.45\textwidth]{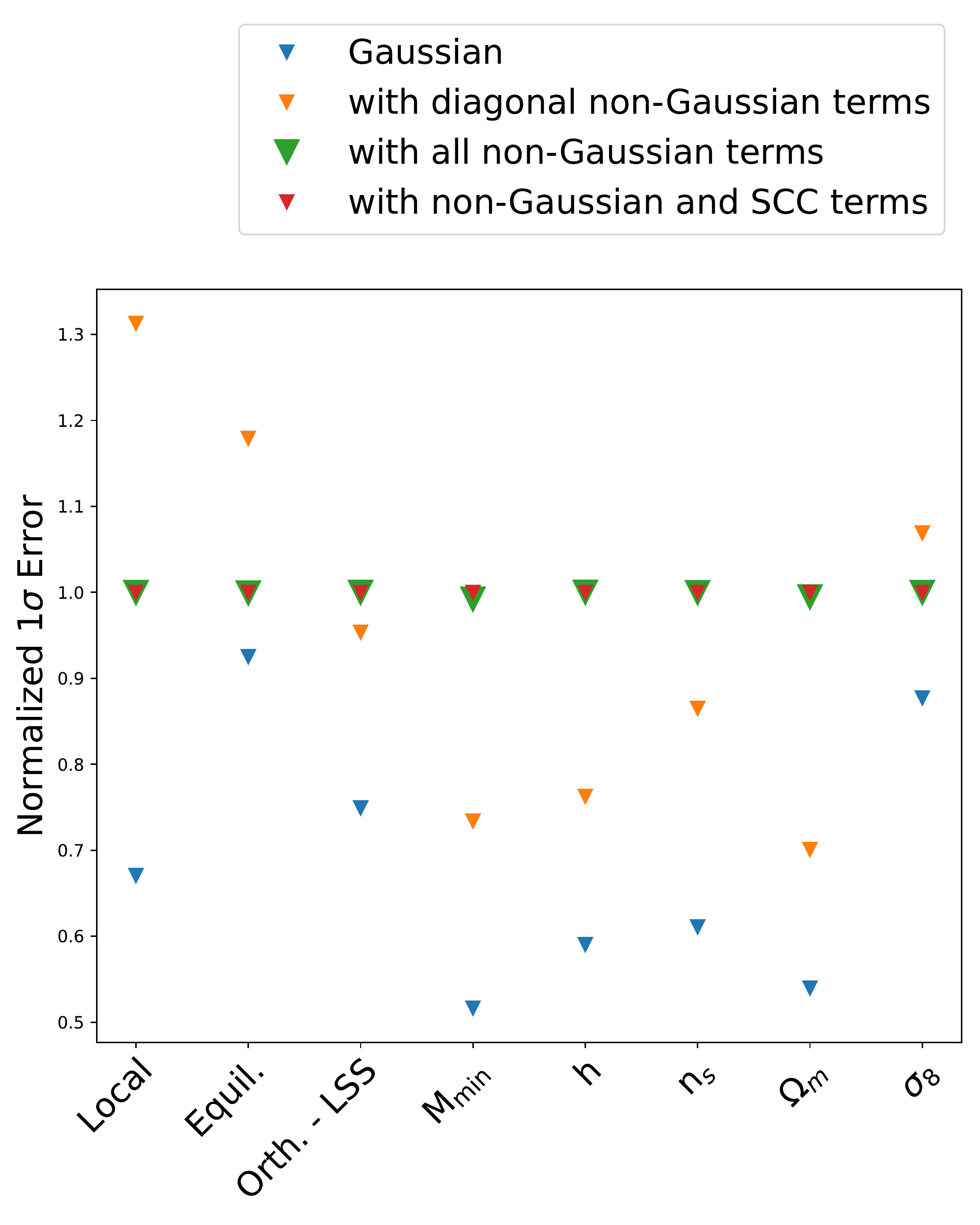}
    \caption{ The impact of various components of the covariance matrix to the joint power spectrum and bispectrum parameter constraints. The constraints are normalized by the error obtained by the full covariance matrix (i.e. including all three terms in Eq. \ref{eq:covMatContributions}). Whilst the non-Gaussian contributions to the covariance matrix are very important for parameter constraints, the super sample covariance terms are not - this is seen as the red and green symbols overlap entirely for most parameters (for clarity we have enlarged the green symbols). \label{fig:cov_types_joint} }
 \end{figure}

The parameter constraints obtained when only considering subsets of the covariance matrix are shown in Fig. \ref{fig:cov_types_joint}. As is expected given the large contribution of the non-Gaussian terms shown in \citet{Chan_2017,Biagetti_2021}, we see that only accounting for the Gaussian terms can lead to a $\sim 100\%$ error on the resulting parameter constraints. Interestingly our results are qualitatively similar to those reported in \citet{dePutter_2018,Floss_2022}, where a perturbative treatment of the leading non-Gaussian terms was performed. Note that this factor is significantly smaller than the equivalent factor for power spectrum and bispectrum measurements of the matter field \citet{Coulton_2022a}, this is because the shot noise contribution to the halo field covariance matrix dominates on small scales and reduces the significance of the non-Gaussian terms. It can also be seen that the off-diagonal terms play an equally important role to the diagonal terms and lead to reductions, as well as increases, in the parameter constraints.  These reductions occur as the power spectrum measurements act as an ancillary statistic and can remove effective noise in the bispectrum - see \citet{Biagetti_2021,Jung_2022} for a more detailed description.

Finally we see that, similar to the matter field, the SSC terms generally only lead to small changes in the parameter constraints. This arises as squeezed bispectrum configurations, which are important for our parameter constraints, are minimally impacted by the SSC terms \citep{Chan_2018,Barreira_2019}. Given the scalings of the SSC term with volume, see e.g. \citet{Li_2014} for a discussion of this, the relative contribution of this term is not expected to become more important for larger volumes. 

\section{Conclusions}\label{sec:Conclusions}
In this paper we have used numerical simulations to quantify the information on primordial non-Gaussainity accessible with measurements of the halo power spectrum, matter power spectrum, halo-matter cross spectrum, and halo bispectrum. For the first time we explore the information content of the halo field including non-linear scales and the full, non-Gaussian covariance matrix. 

First, for the sample considered here, we find that the scale dependent bias feature contains almost all of the information on \emph{local} PNG accessible with power spectrum and bispectrum measurements. Thus, when using sample variance cancellation techniques, only measurements of the large scale power spectrum are needed - especially as there is minimal degeneracy with the other parameters considered here. These results reinforce the analytical results of \citet{dePutter_2018} who found that, for tracer number densities below $\bar{n}\sim \mathrm{few} \times 10^{-3}\mathrm{h}^3/\mathrm{Mpc}^{3}$, the bispectrum and trispectrum contains minimal additional information. Interestingly, when compared to the primordial field, the scale dependent bias feature only contains information equivalent to $k\approx 0.15~ \mathrm{h/Mpc}$, indicating that alternative late-time statistical probes \citep[such as those discussed in e.g.][]{Uhlemann_2018,Friedrich_2020,Cole_2020,Biagetti_2022,Giri_2022,Andrews_2022}, may be able to provide additional information.

Second, for \emph{equilateral} and \emph{orthogonal-LSS} PNG, which have very weak scale dependent bias features, we find that bispectrum measurements provide more constraining power than the power spectrum. However both shapes exhibit strong degeneracies with the cosmological and bias parameters, especially \emph{equilateral}. For the \emph{orthogonal-LSS} shape, these degeneracies are largely broken as smaller scale information is included or through combining the bispectrum measurements with the power spectrum measurements. However this is not the case for the \emph{equilateral} shape, highlighting the challenge of measuring this shape with bispectrum measurements. 

Similar to the results found for the matter bispectrum \citep{Coulton_2022a,Jung_2022}, we find that for the \emph{local} and \emph{orthogonal-LSS} shapes the information in the bispectrum seems to saturate at $k_\mathrm{max}\approx 0.35 \mathrm{h/Mpc}$. For the matter bispectrum this occurred due the large non-Gaussian contributions to the covariance matrix. For the halo bispectrum it occurs due to the contribution from shot-noise, which dominates on small scales. This suggests that more information may be accessible in the halo field, and therefore the galaxy field, if the shot noise were suppressed (e.g. for a higher density tracer). Additionally we find that the super sample covariance contribution to the covariance is negligible, generalizing the results of the squeezed limit bispectrum found in \citet{Barreira_2019}.

The key limitation of our \emph{local} PNG analysis is that we did not consider the impact of marginalizing over $b_\phi$, Eq. \ref{eq:haloMatterBias}, which is important as the constraints from scale dependent bias are sensitive only to the product $b_\phi f_\mathrm{NL}$. Our constraints are entirely driven by the scale dependent bias effect and we would not be able to break the degeneracy between $b_\phi$ and $f_\mathrm{NL}$ with the bispectrum measurements. Thus it would be conservative to interpret our results as constraints for $b_\phi f_\mathrm{NL}$. Whilst this degeneracy would not impact our ability to detect the existence of \emph{local} PNG, it would complicate any interpretation of the specific value. The specific value would rely on the assumptions used in modelling or maginalizing $b_\phi$, which depend upon the properties of the specific tracer. A trivial demonstration of this point was seen for \emph{orthogonal-CMB} PNG - this type of PNG produces a $1/k$ scale dependent bias but for the tracer considered here the associated bias coefficient is close to zero. This is a trivial case as for our halos the bias parameter can be computed with reasonable accuracy, however it highlights how small bias coefficients can mask the PNG signals. \citet{Moradinezhad_2021} explored how the choice of priors in marginalizing over $b_\phi$ impact constraints and how the addition of bispectrum measurements affect this degeneracy. Informative $b_\phi$ priors allow strong $f_\mathrm{NL}$ inferences, but care must be taken to avoid biased results \citep{Moradinezhad_2021} . This issue is discussed in more depth in \citet{Barreira_2022b}.  In this work, we have also neglected relativistic effects, which can introduce similar features \citep{Bruni_2012,Jeong_2012,Camera_2015,Castorina_2022} though these can be differentiated via the different redshift evolution.

Similarly an important and substantial limitation of our \emph{equilateral} and \emph{orthogonal} analysis is our bias model. Our bias model effectively only includes the leading bias term and neglects all higher order biases, which is a gross simplification. We also investigated two variations on our bias models: first we considered replacing the halo mass cut-off, $M_{min}$, with a linear bias parameter $b_1$. We found that using $b_1$ leads to very similar constraints to using the halo bias parameter, which arises as the two impact our statistics, in the ranges relevant for our bispectrum constraint, in a similar manner . The second test involved fitting constraints without any bias modelling. When we fit all the parameters jointly the contours only marginally widened when including the bias parameters, largely due to the large degeneracies with cosmological parameters discussed above. However if we fix the cosmological parameters including the bias parameter doubles the equilateral constraint, whilst leaving the local constraint largely unchanged. The constraints presented here would likely by further degraded when marginalizing over a set of realistic bias uncertainties, especially tidal biases, thus we view our results as `best case' estimate. In future work we will analyze this in more detail, by for example using a halo occupation distribution (HOD) to marginalize over more realistic biases as was considered in \citet{Hahn_2021}. Similarly it would be highly interesting to consider different halo mass samples. Unfortunately the small number of objects present in our simulations means that we cannot further divide our halo catalog without increasing the convergence issues faced in this analysis. 

A technical challenge of this work arose due to the large shot-noise found in the halo bispectrum, which makes it difficult to accurately measure the bispectrum derivatives. This issue is examined in detail in Appendix \ref{app:derivConvergence} where we ensure that our results were robust by implementing two methods to mitigate the impact of this noise. First through removing the noise by smoothing with a Gaussian process and second by using a newly developed method to estimate the Fisher information \citep{Coulton_2022b}. Without including these mitigation methods our forecast constraints would be biased too small by a factor of $>2$! This difficulty highlights the challenges that shot-noise from halos (and galaxies) poses for analyses exploiting small scale information and, more generally, the need to carefully examine the convergence of numerical derivatives.

\begin{acknowledgments}
The authors are very grateful to Simone Ferraro, Oliver Philcox, Colin Hill, Daan Meerburg, Shirley Ho, David Spergel and Yin Li for useful discussions throughout this work. 
GJ, ML and MB were supported by the project "Combining Cosmic Microwave Background and Large Scale Structure data: an Integrated Approach for Addressing Fundamental Questions in Cosmology", funded by the MIUR Progetti di Ricerca di Rilevante Interesse Nazionale (PRIN) Bando 2017 - grant 2017YJYZAH.
DK is supported by the South African Radio Astronomy Observatory (SARAO)
and the National Research Foundation (Grant No. 75415). B.D.W. acknowledges support by the ANR BIG4 project, grant ANR-16-CE23-0002 of the French Agence Nationale de la Recherche; and the Labex ILP (reference ANR-10-LABX-63) part of the Idex SUPER, and received financial state aid managed by the Agence Nationale de la Recherche, as part of the programme Investissements d'avenir under the reference ANR-11-IDEX-0004-02.
The Flatiron Institute is supported by the Simons Foundation.
LV acknowledges  ERC (BePreSySe, grant agree- ment 725327),  PGC2018-098866- B-I00 MCIN/AEI/10.13039/501100011033 y FEDER “Una manera de hacer Europa”, and the “Center of Excellence Maria de Maeztu 2020-2023” award to the ICCUB (CEX2019-000918-M funded by MCIN/AEI/10.13039/501100011033).
\end{acknowledgments}

\bibliographystyle{apsrev.bst}
\bibliography{png,planck_bib}

\appendix
\section{Derivative convergence analysis}\label{app:derivConvergence}
\begin{figure*}
  \includegraphics[width=1.05\textwidth]{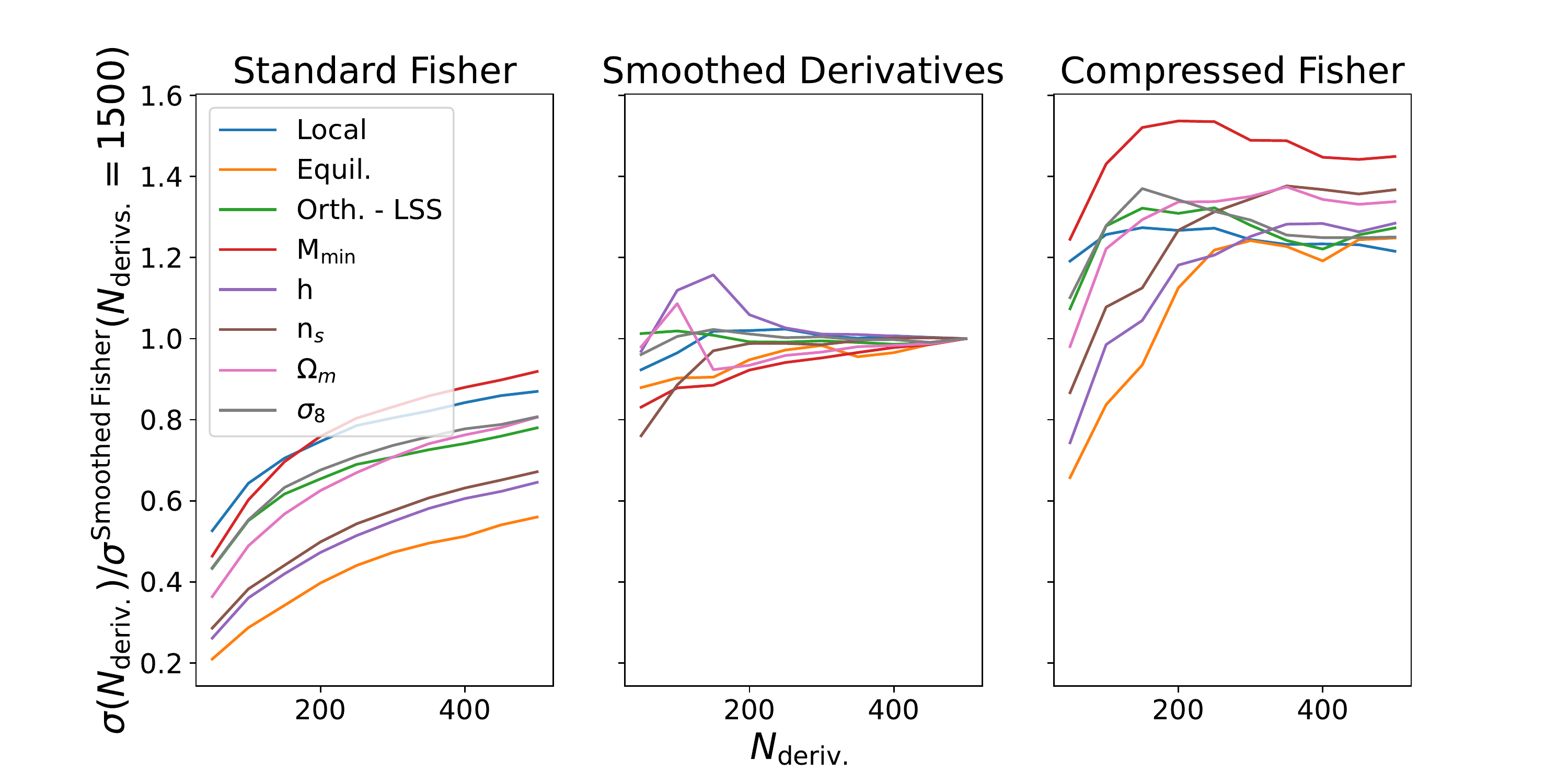}
    \caption{ A examination of the stability of the Fisher forecast to changes in the number of simulations. In the left plot we plot the results of the standard Fisher forecast. In the center plot we show the results when we use a Gaussian Process to mitigate noise in the derivatives. In the right plot we show an alternative Fisher forecast method using data compression. This last method, for $N_\mathrm{deriv}\gtrsim 250$, provides an approximate upper bound on the Fisher information. To easily compare across the different parameters, the Fisher estimates are normalized by the GP estimate for each parameter.   \label{fig:convergnece_bispec} }
 \end{figure*}

One method of verifying the convergence is to investigate whether the resulting Fisher information, and estimated parameter constraints - $\sigma^2_i = F^{-1}_{ii}$, are stable to variations in the number of simulations used. In Fig. \ref{fig:convergnece_bispec} we explore the stability of the bispectrum constraints. We find that the constraints show large variations as the number of simulations is varied and the constraints scale roughly like the square root of the number of derivative simulations! This implies that the resulting constraints are not converged and not reliable. 

The reason for this lack of convergence is the large noise in the derivatives, visible in Fig \ref{fig:bispec_wGP}.We can understand this behaviour by computing the expectation of the standard Fisher information estimator 
\begin{align}
    \langle \hat{F_{ij}}\rangle = \frac{\mathrm{d}\mu}{\mathrm{d}\theta_i} \Sigma^{-1}\frac{\mathrm{d}\mu}{\mathrm{d}\theta_j}+ \Sigma^{-1} Cov\left[\frac{\mathrm{d}\mu}{\mathrm{d}\theta_i},\frac{\mathrm{d}\mu}{\mathrm{d}\theta_j} \right],
\end{align}
where we have ignored any noise in the covariance matrix. We see that the Fisher information is biased high due to the covariance of the mean of the derivatives. Hence the parameter constraints, which depend on the inverse, are biased low. The covariance in the derivatives decreases as we utilze more simulations and hence the bias decreases monotonically as the number of simulations increases. Note that previous works investigating the information content of the \textsc{quijote} halo bispectrum likely also suffer this issue and we present a short discussion of this in Appendix \ref{app:mnu_constraints} .

To generate robust constraints we need to mitigate this noise and in the following sections we discuss two such methods.

\subsection{Smoothed derivatives}\label{app:deriv_smoothing}
One method of removing the impact of noise would be to fit a function through the derivatives and use this in the Fisher forecast. By eye it is suggestive that, for most of the configurations and PNG shapes seen in  Fig \ref{fig:bispec_wGP}, the measured derivatives could be described by a smooth function. Unfortunately it is challenging to write down a function that captures the full shape over all the scales (perturbative models are not able to capture the range of scales considered here). 

As an alternative approach we explored using a Gaussian process (GP) to perform a non-parametric estimation of the underlying function, see \citet{RasmussenW06} for a review of Gaussian processes.  In this approach we treat the measured derivatives as coming from a stochastic process, in this case a zero mean Gaussian process, where we additionally use the measured variance of each point to assign a measurement uncertainty to each point. The intuition with this approach is the following: instead of specifying a full functional form, as in parametric approaches, we ask given a set of data points what can I infer about the value of the function at another value. In this stochastic process framework the relation of the data at one point to any another point (observed or not) is given by the correlation structure of the process thus information is contained in the structure of the Gaussian process covariance matrix - i.e. the structure of the function is implicitly determined by the data points and the correlations amongst them. Given our data points (the derivatives form the simulations) and their noise, we fit for the properties of this Gaussian covariance matrix. The noise assigned to each point is vital as it allows the Gaussian process to smooth out the noise - it estimates what smooth functions are consistent with the data points, given their measurement noise.  Note that we could have fit the Gaussian process directly to the simulations at $\theta$,$\theta+\delta\theta$ and $\theta-\delta \theta$ including dependence on the parameters in the Gaussian process. Then we can use the Gaussian process to estimate the derivatives. Empirically it was found that fitting the Gaussian processes to the derivatives directly worked better. We attribute this difference to sample variance cancellation - our simulations have matched initial seeds so when we compute the derivatives by central differencing the majority of the sample variance cancels. We fitting the GP directly to the simulations at $\theta$,$\theta+\delta\theta$ and $\theta-\delta \theta$ it is difficult to incorporate this sample variance cancellation information and without it we found the GP was not able to accurately model the derivatives.

We use the implementation provided in the scikit-learn library \citep{scikit-learn}. We use an isotropic `Radial-basis function' kernel, i.e. a Gaussian kernel, to describe the correlations between data points. The parameters characterizing the Gaussian process are estimated by maximizing the log-marginal-likelihood from the derivatives. Before fitting the Gaussian process to the derivatives we divide the data by a smooth function to reduce the dynamic range of the problem. This provided significantly better performance. For the $\Lambda$CDM parameters we divide the derivatives by the mean of the covariance matrix simulations. For the PNG derivatives, we use a linear combination of the tree level PNG bispectra and the mean of the covariance matrix (this combination helps capture the general behaviour on both the large and small scales). The measurement error assigned to each point when fitting the GP is the error on the mean of our simulated derivatives. Once the derivatives have been smoothed by our Gaussian process we multiply back by the same base functions.

In Fig. \ref{fig:bispec_wGP} we compare the resulting smoothed bispectra to the measurements. There are several notable features: firstly where the bispectra are well measured the GP matches the measurements very well. In the highly noisy region the GP is a significantly smoother function producing a curve that is consistent with the simulation points, given their errors. In a few places the GP shows statistically significant deviations from the data points - this demonstrates the limitations of this method. These points typically occur when there is a steep gradient in one direction of the bispectrum and thus these points represent an over smoothing. Despite this caveat, the smoothed derivatives generally provide a good, smoothed approximation of the simulations.

Given this smoothing method, we can investigate how our constraining power changes as the number of simulations used to compute the smoothed derivatives is varied. This test allows us to assess the effectiveness of our smoothing procedure: if the derivatives are not sufficiently smoothed we expected to see constraints that increase as more simulations are added; on the other hand, if the smoothing process is too aggressive, and is removing important physical features, we may see constraints that decrease as the number of simulations are varied. As is seen in Fig. \ref{fig:convergnece_bispec}, our constraints are now stable to variations in the number of simulations, thus providing a useful validation of our smoothing method.

\subsection{Compressed Fisher Forecast}
Given the potential caveats of the smoothed approach, it is worthwhile to explore alternative methods of validating our Fisher forecast. Recent work by \citet{Coulton_2022b} has demonstrated an alternative method of performing Fisher forecasts that, under certain conditions, leads to conservative forecast constraints. Thus this method allows another test of our smoothed derivatives and can be combined with the standard Fisher method to provide bounds on the constraints.  Here we briefly overview the method, hereafter referred to as the Compressed Fisher forecast, and refer the reader to \citet{Coulton_2022b} for more details.

In this approach instead of computing the Fisher information of the power spectrum and bispectrum, we compute the Fisher information of a set of compressed summary statistics. In principle the information in the power spectrum and bispectrum can be losslessly compressed to a set of statistics with the same dimension as the number of parameters (this includes both cosmological parameters and nuisance parameters). One method of achieving this is to compress the data via \citep{Alsing_2018}
\begin{align}\label{eq:compression_general}
\mathbf{t} = \nabla \mathcal{L}(\mathbf{d},\theta),
\end{align}
where $\mathbf{t}$ are the summary statistics, $ \mathcal{L}(\mathbf{d},\theta)$ is the likelihood for data, $\mathbf{d}$, given parameters, $\theta$, and the derivative is with respect to the parameters of interest. In this compressed space, the Fisher information is
\begin{align}\label{eq:compressed_fisher}
F^{tt}_{ij} = \frac{\partial \mathbf{\mu}^t } {\partial \theta_i} {\Sigma^{tt}}^{-1}\frac{\partial \mathbf{\mu}^t } {\partial \theta_j}, 
\end{align}
where $\mu^t$ and $\Sigma^{tt}$ are the mean and covariance of the compressed statistics.
Performing the Fisher analysis in the compressed space has two advantages: first, the dimension of the data vectors are drastically reduced so a smaller dimension covariance matrix is needed. Second, the noise on the derivatives of the compressed statistic, which is required for the Fisher forecast, is expected to be lower as it is a weighted sum of all the data points and thereby averages down the noise in the uncompressed derivatives.

For the case of a Gaussian likelihood with parameter independent covariance, as is used here and motivated by \citet{Scoccimarro_2000,Carron_2013}, the lossless compression is \citep{Tegmark_1997,Heavens_2000}
\begin{align}
\mathbf{t} = \frac{\partial \mathbf{\mu}}{\partial \theta} C^{-1} \left( \mathbf{d} - \mathbf{\mu} \right),
\end{align}
where $\mu$ and $C$ are the data mean and covariance. Compression has already been consider for bispectrum analysis \citep[e.g.][]{Gualdi_2019,Philcox_2021,Byun_2022} to aid working with the very high dimensional data vectors, but here we are using it to improve the accuracy of the Fisher forecast (and we use a simulation-based rather than perturbative model). An issue that is immediately clear is that performing the optimal compression requires the same ingredients as performing the standard Fisher forecast. This issue can resolved by splitting the simulations into two parts: the first part is used to estimate the compression and the second part is then compressed and used to estimate the quantities needed for the compressed Fisher forecast, Eq. \ref{eq:compressed_fisher}. This splitting is necessary as reusing simulations will lead to biased Fisher estimates. In this work we use 75\% of the data to estimate the compression and 25\% to compute the compressed Fisher information. The use of estimated quantities in the compression results in a suboptimal compression and so the resulting Fisher forecast errors will be larger than the truth \citep[see e.g.][]{LehmCase98}. Thus we have traded forecasts errors that were biased too small to a set of conservative errors that are bias to be too large. In the limit of a sufficient number of simulations the standard Fisher and compressed Fisher estimates tend to the same values. Note that when the noise in the compressed derivatives is large, which occurs for small number of simulations, this estimator will be biased low in the same manner as the standard estimator.

\begin{figure*}
  \includegraphics[width=1.0\textwidth]{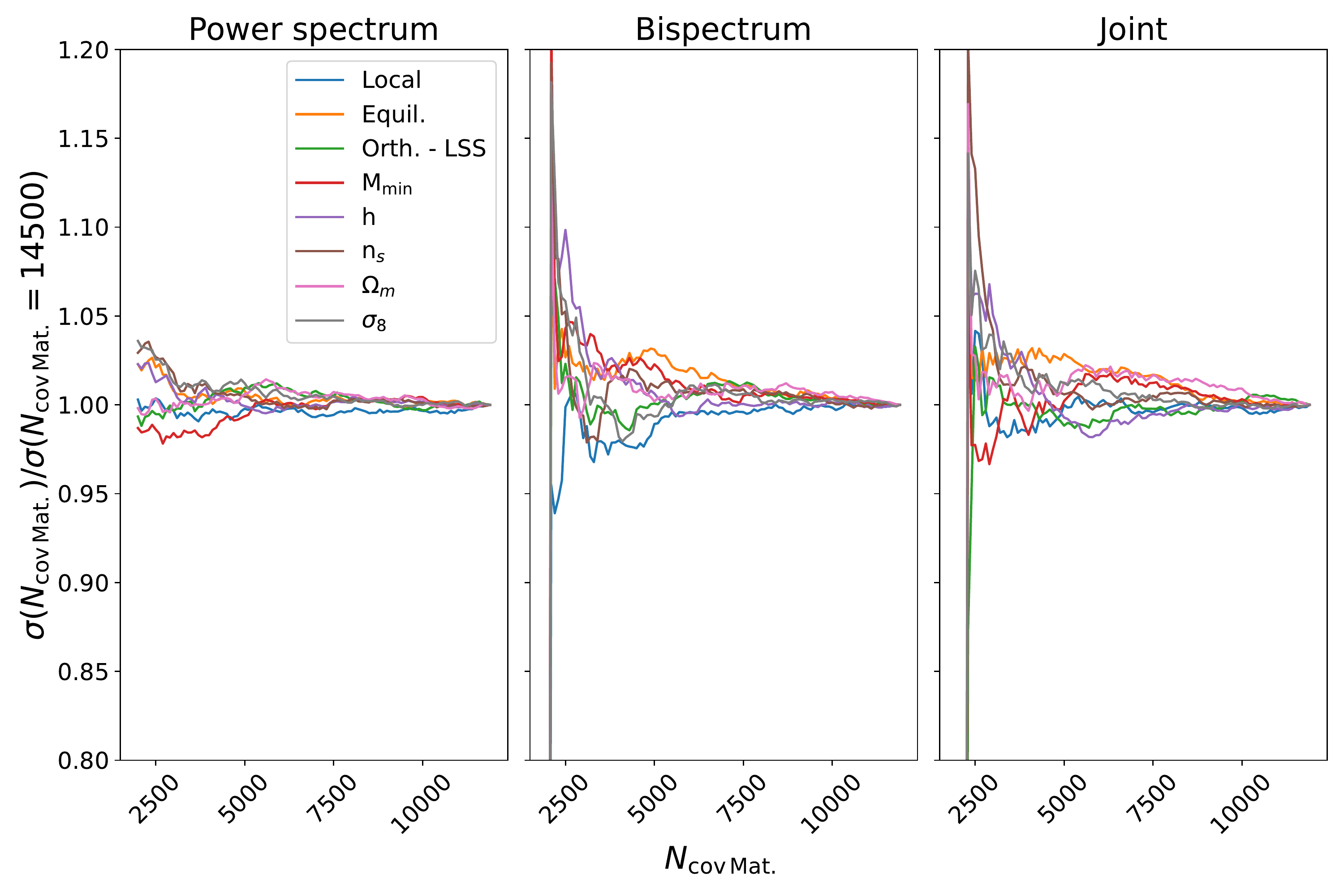}
    \caption{ The stability of our joint constraints to variations in the number of simulations used to compute the covariance matrix. The small variations seen demonstrate that the covariance matrix used in our full analysis, computed with 14,500 simulations, is converged. \label{fig:covMatConvergence} }
 \end{figure*}
In Fig. \ref{fig:convergnece_bispec} we investigate the stability of our compressed analysis to the number of simulations used to compute the derivatives. Unlike the standard estimate, the compressed forecast reaches a regime where it is stable to changes in the number of simulations used: for most of the parameters the forecast errors are largely unchanged or they decrease as more than 250 simulations are used. The decrease occurs as, once convergence is reached, using more simulations only improves the effectiveness of the compression. Note that for very small numbers of simulations, $\lesssim 250$, the forecast errors based on the compressed Fisher method is biased low due to noisy estimates of the derivatives, in an identical effect to the standard forecast. See \citet{Coulton_2022b} for a detailed discussion of this estimator, as well as an improved estimator that accelerates convergence to the exact Fisher forecast.

Given that our compressed Fisher forecast is converged, we can compare these forecast constraints to those of the other methods. In Fig. \ref{fig:convergnece_bispec} we see that, as expected, the forecast constraints are larger than the GP constraints and the standard (and unconverged) forecast. This hierarchy provides a conservative estimate of the information available and validation that our smoothing method does not wash out the important features.
 
\section{Convergence of the covariance matrix}\label{app:covMat_convergences}
In Fig. \ref{fig:covMatConvergence} we demonstrate that our forecast constraints only exhibit a $\leq 2\%$ variation as the number of simulations is changed from 7000 to 14500 and thus demonstrates that our analysis is likely converged with respect to the number of covariance matrix simulations.

\section{Neutrino Mass Constraints}\label{app:mnu_constraints}
In the course of developing our results, we reexamined the power of halo bispectrum measurements in constraining neutrino mass. Neutrino mass constraints are highly challenging as the physical impact of observationally relevant neutrino masses is very small. We find that with the number of simulations used in \citet{Hahn_2020} the bispectrum derivatives are not converged. Given this we apply our Gaussian process smoothing method, as described in Appendix \ref{app:deriv_smoothing}, to this analysis. Note that we do not fit $b_1$ and $\Omega_b$ here as was done in  \citet{Hahn_2020} to ensure converged results.

\begin{figure*}
  \includegraphics[width=1.\textwidth]{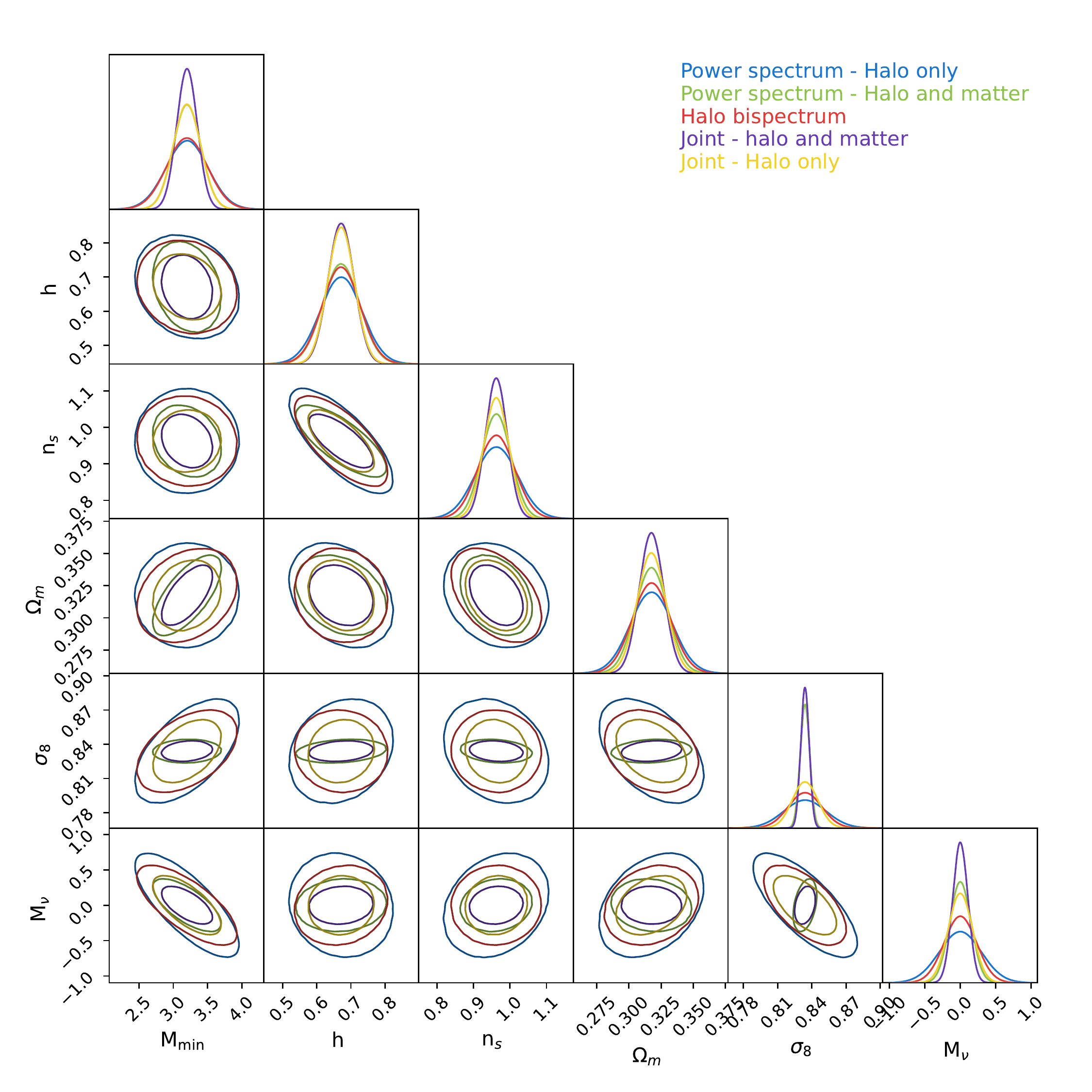}
    \caption{ Similar to Fig. \ref{fig:param_const_corner} however for constraints on neutrino mass. Using the Gaussian process smoothed derivatives, which are required for converged forecasts with our simulations, the bispectrum measurements alone have minimal constraining power on neutrino mass.\label{fig:param_const_corner_mnu} } 
 \end{figure*}

In Fig. \ref{fig:param_const_corner_mnu} we show the new constraints, finding that the halo-bispectrum constraints are dramatically degraded. On its own we find that the halo bispectrum has slightly more constraining power than the power spectrum on the sum of the masses of the neutrinos (with constraints 1.3 $\times$ smaller compared to 5 $\times$ smaller found in \citet{Hahn_2020}). By combining it with power spectrum measurements - the bispectrum measurements help better constrain the $\Lambda$CDM and bias parameters and break degeneracies, thereby allowing further improved constraints.

\end{document}